\documentclass[aps,twocolumn,showpacs,preprintnumbers,amsmath,amssymb,nofootinbib,showkeys]{revtex4-1}

\RequirePackage[T1]{fontenc}

\usepackage{epsfig,graphicx,amssymb,amsmath}
\RequirePackage{mathptmx}      
\RequirePackage{flushend}
\RequirePackage{color}
\usepackage{braket}
\usepackage{changes}
\usepackage{multirow}
\RequirePackage{hyperref}
\hypersetup{
	linktocpage,
    colorlinks,
    citecolor=blue,
    filecolor=black,
    linkcolor=blue,
    urlcolor=blue,
}

\begin{document}

\title{Sensitivity of the $\eta^{(\prime)}\to\pi^{0}\gamma\gamma$ and 
$\eta^{\prime}\to\eta\gamma\gamma$ decays to a sub-GeV leptophobic $U(1)_{B}$ boson}


\author{Rafel Escribano$^{1,2}$}\email{rescriba@ifae.es}
\author{Sergi Gonz\`{a}lez-Sol\'{i}s$^{3}$}\email{sergig@lanl.gov}
\author{Emilio Royo$^{1,2}$} \email{eroyo@ifae.es}

\affiliation{$^1$Grup de F\'{i}sica Te\`{o}rica, Departament de F\'{i}sica, 
Universitat Aut\`{o}noma de Barcelona, 08193 Bellaterra, Barcelona, Spain\\
$^2$Institut de F\'{i}sica d'Altes Energies (IFAE) and 
The Barcelona Institute of Science and Technology, Campus UAB, 08193 Bellaterra, Barcelona, Spain\\
$^3$Theoretical Division, Los Alamos National Laboratory, Los Alamos, New Mexico 87545, USA}


\preprint{LA-UR-22-26889}


\begin{abstract}

The sensitivity of the rare decays $\eta^{(\prime)}\to\pi^{0}\gamma\gamma$ and 
$\eta^{\prime}\to\eta\gamma\gamma$ to signatures of a leptophobic $B$ boson in the MeV--GeV mass range 
is analyzed in this work.
By adding an explicit $B$-boson resonance exchange, $\eta\to B\gamma\to\pi^{0}\gamma\gamma$, 
to the Standard Model contributions from 
vector and scalar meson exchanges, 
and employing experimental data for the associated branching ratios, 
it allows us to improve the current constraints on the $B$-boson mass $m_{B}$ 
and coupling to Standard Model particles $\alpha_{B}$.
From these constraints and the analysis of the available experimental $\gamma\gamma$ 
invariant mass distribution, we show that a $B$-boson signature in the resonant mass range 
$m_{\pi^{0}}\lesssim m_{B}\lesssim m_{\eta}$ is strongly suppressed and would be very difficult to 
experimentally identify, assuming that the leptophobic $B$ boson only decays to Standard Model particles.
In contrast, the limits outside this mass window are less stringent and the corresponding $t$- and 
$u$-channel signatures may still be observable in the data, as it occurs with the nonresonant 
Standard Model $\rho$, $\omega$ and $\phi$ meson exchanges. 
In addition, we make use of experimental data from the $\eta^{\prime}\to\pi^{0}\gamma\gamma$ and 
$\eta^{\prime}\to\eta\gamma\gamma$ decays to explore larger $B$-boson masses.
Our results are relevant for the $B$-boson search programs at existing and forthcoming 
light-meson facilities, such as KLOE(-II) and Jefferson Lab Eta Factory experiments.


\end{abstract}


\maketitle


\section{Introduction}\label{section1}

An increasingly ubiquitous strategy to search for physical phenomena beyond the Standard Model (BSM) 
is to test fundamental symmetries such as $C$, $P$, $T$, $CP$, and $CPT$ in different processes.
Specifically, decays of the neutral pseudoscalar mesons $\eta$ and $\eta^{\prime}$ 
constitute a particularly suited playground to look for new physics~\cite{Gan:2020aco,REDTOP:2022slw}. 
This is because these two mesons are special, 
as they are eigenstates of the $C$, $P$, $CP$ and $G$-parity operators, 
and all their strong and electromagnetic decays are either anomalous or forbidden at lowest order 
due to the conservation of fundamental symmetries of QCD. 
Consequently, higher-order contributions are expected to become relevant, 
rendering the $\eta/\eta^{\prime}$ decays sensitive hadronic probes to test discrete symmetries 
and to search for undiscovered fundamental BSM particles, 
such as dark photons or leptophobic $U(1)_{B}$ bosons 
(see Refs.~\cite{Gan:2020aco,REDTOP:2022slw} and references therein).
Examples of this are the rare $\eta^{(\prime)}\to\pi^{0}\gamma\gamma$ and 
$\eta^{\prime}\to\eta\gamma\gamma$ decays which, 
as they are highly suppressed in the 
Standard Model (SM)~\cite{Ametller:1991dp,Oset:2002sh,Oset:2008hp,Achasov:2001qm,Escribano:2018cwg}, 
have been put forward as fine probes to search for MeV--GeV signatures of a new leptophobic 
$B$ boson~\cite{Tulin:2014tya} arising from a new $U(1)_{B}$ gauge symmetry 
which couples predominantly to quarks over leptons~\cite{Lee:1955vk,Nelson:1989fx,Rajpoot:1989jb,Foot:1989ts,He:1989mi}.

Experimental searches for leptophobic $B$ bosons depend on the mass $m_{B}$ and the associated decay channels, and have placed constraints on the coupling for masses that span from below the MeV scale, obtained from long-range nuclear forces~\cite{Adelberger:2003zx} and low-energy neutron scattering~\cite{Barbieri:1975xy,Leeb:1992qf,Nesvizhevsky:2007by}, to above the GeV scale, obtained at high-energy hadron colliders in dijet resonance searches, as well as in heavy quarkonia and $Z$ decays~\cite{Carone:1994aa,Bailey:1994qv,Carone:1995pu,Aranda:1998fr,Dobrescu:2013cmh,Dobrescu:2014fca}.
The intermediate MeV--GeV mass range has been less explored thus far~\cite{Nelson:1989fx,Tulin:2014tya}, which is down to this being the region of nonperturbative QCD, and has often been considered as a challenging blindspot for experiment in the past.
However, searches for leptophobic $B$ bosons are gaining traction in this intermediate mass range given the potential signatures that can be looked for in decays of light mesons, such as $\eta$, $\eta^{\prime}$, $\omega$, and $\phi$~\cite{Tulin:2014tya}, after years of sterile sub-GeV dark-photon searches most of them relying on the coupling of this new force to leptons in decays to $e^{+}e^{-}$ and $\mu^{+}\mu^{-}$ pairs~\cite{APEX:2011dww,Essig:2013lka,NA482:2015wmo,Alexander:2016aln,KLOE-2:2012lii,Anastasi:2015qla,KLOE-2:2018kqf,BaBar:2014zli,Merkel:2014avp,HPS:2018xkw,LHCb:2019vmc}.
In fact, the search for leptophobic $B$ bosons has been incorporated into the physics programs of existing light-meson factories such as the KLOE-II Collaboration, which is searching for $B$ bosons by looking for enhancements in the $\pi^{0}\gamma$ invariant mass spectrum of the $\phi\to\eta B\to\eta\pi^{0}\gamma$ process~\cite{delRio:2021xag,Cao:2022rxo}, and is a top priority physics goal for the recently approved Jefferson Lab Eta Factory (JEF) experiment~\cite{JEF}, which promises a new and exciting era for $\eta$ and $\eta^{\prime}$ physics, with the $\eta\to\pi^{0}\gamma\gamma$ decay being their key signal channel.
The Belle Collaboration has also pursued searches for a $B$ boson decaying into $\pi^{+}\pi^{-}$ using $\eta\to\pi^{+}\pi^{-}\gamma$ decays~\cite{Belle:2016uaq} but have found no signal and, since $B\to\pi^{+}\pi^{-}$ is suppressed by $G$-parity conservation, the limits on the $B$-boson parameters are not as stringent as the constraints coming from other decays, such as the $\eta\to\pi^{0}\gamma\gamma$.
$B$-boson searches may also be carried out at future $\eta/\eta^{\prime}$ factories, such as the proposed REDTOP experiment~\cite{REDTOP:2022slw},\footnote{The current detector layout at REDTOP is nonsensitive to 
neutral final states~\cite{REDTOP:2022slw}, but an improved version of REDTOP is planned where the $\eta$ will be tagged and final states with $\pi^{0}$'s and photons could be detected.} or in direct photoproduction $\gamma p\to Bp\to\pi^{+}\pi^{-}\pi^{0}p$ at the GlueX experiment at Jefferson Lab~\cite{Fanelli:2016utb}, which will probe $B$-boson masses above 0.5 GeV.
Alternatively, signals of $U(1)_{B}$ leptophobic $B$ bosons decaying into invisible particles, i.e.~dark matter, have also been pursued at neutrino factories~\cite{Batell:2014yra} and at the LHC~\cite{Cohen:2015toa}. 

The model that we consider in this work for a $U(1)_{B}$ leptophobic gauge boson $B$ that couples to the baryon number has the following interaction Lagrangian~\cite{Nelson:1989fx,Tulin:2014tya}
\begin{equation}
\mathcal{L}_{\rm{int}}=
\left(\frac{1}{3}g_{B}+\varepsilon Q_{q}e\right)\bar{q}\gamma^{\mu}qB_{\mu}
-\varepsilon e\bar{\ell}\gamma^{\mu}\ell B_{\mu}\ ,   
\label{Eq:BbosonLagrangian}
\end{equation}
where $B_{\mu}$ is the new gauge boson field and $g_{B}$ is the new gauge coupling, 
with $\alpha_{B}=g_{B}^{2}/4\pi$ being the fine structure constant associated to the baryonic force.
This interaction structure is gauge invariant and preserves the low-energy symmetries of QCD, 
namely $C$, $P$ and $T$ invariance, as well as isospin and $SU(3)$-flavor symmetry.

Partial widths for $B$-boson decays in the MeV--GeV mass range have been calculated in~\cite{Tulin:2014tya} using the hidden local symmetry framework for vector meson dominance (VMD).
Above the single-pion threshold, $m_{\pi^{0}}\lesssim m_{B}\lesssim1$ GeV, the $B$ boson decays predominantly to $\pi^{0}\gamma$, or to $\pi^{0}\pi^{+}\pi^{-}$ when kinematically allowed, very much like the $\omega$ meson. In fact, the $B$ boson can be assigned the same quantum numbers as those from the $\omega$, i.e.~$I^{G}(J^{PC})=0^{-}(1^{--})$.
It must be noted that the interaction Lagrangian in Eq.~(\ref{Eq:BbosonLagrangian}) is not completely decoupled from leptons as it contains subleading photonlike couplings to leptons proportional to $\varepsilon=eg_{B}/(4\pi)^{2}$.
This effect allows the purely leptonic decay $B\to e^{+}e^{-}$, which dominates below single-pion threshold $m_{B}\lesssim m_{\pi^{0}}$.
There are other allowed decay channels such as $B\to\eta\gamma$ and $B\to\pi^{+}\pi^{-}$; however, these are subleading~\cite{Tulin:2014tya}, where the latter, being forbidden by $G$-parity conservation, arises via $\rho$-$\omega$ mixing. 

At present, conservative constraints from $\eta$ and $\eta^{\prime}$ decays on the $B$-boson parameters $\alpha_{B}$ and $m_{B}$ are based on total rates setting the SM contribution to zero~\cite{Tulin:2014tya}, i.e.~assuming the $B$-boson intermediate states $\eta^{(\prime)}\to B\gamma\to\pi^{0}\gamma\gamma$, $\eta^{\prime}\to B\gamma\to\pi^{+}\pi^{-}\pi^{0}\gamma$ and $\eta^{\prime}\to B\gamma\to\eta\gamma\gamma$, and making use of the narrow width approximation (NWA), e.g.~$\mbox{BR}(\eta\to\pi^{0}\gamma\gamma)=\mbox{BR}(\eta\to B\gamma)\times\mbox{BR}(B\to\pi^{0}\gamma)$.
It must be stressed, though, that the SM contribution to these decays is not negligible~\cite{Oset:2008hp,Escribano:2018cwg,Danilkin:2017lyn,Lu:2020qeo} and, therefore, it should not be disregarded in exclusion analyses of $B$ bosons. Thus, one of the goals of the present work is to take into account SM effects in these analyses. 
To that effect, we employ our controlled SM contributions, i.e.~the VMD and linear sigma model (L$\sigma$M) amplitudes from Ref.~\cite{Escribano:2018cwg}, we supplement it with the explicit inclusion of an intermediate $B$ boson and use the most up-to-date experimental data. 

Significantly greater sensitivity to the $B$-boson model could be obtained from the analysis of the invariant mass distributions.
Provided that $m_{\pi^{0}}\leq m_{B}\leq m_{\eta^{(\prime)}}$, the $B$-boson mediated decay $\eta^{(\prime)}\to B\gamma \to\pi^{0}\gamma\gamma$ would reveal a peak at around $m_{B}$ in the $\pi^{0}\gamma$ invariant mass spectrum. 
Searches for a $\pi^{0}\gamma$ resonance within this mass region in $\eta\to\pi^{0}\gamma\gamma$ decays are the main physics goal of the JEF experiment~\cite{JEF}, which plans to improve the total rate limit by two orders of magnitude, and is being searched for by KLOE-II via $\phi\to\eta B\to\eta\pi^{0}\gamma$~\cite{delRio:2021xag,Cao:2022rxo} and $\eta\to B\gamma\to\pi^{0}\gamma\gamma$.
Accordingly, we aim to perform a detailed analysis of the $\gamma\gamma$ and $\pi^{0}\gamma$ invariant mass distributions. 
In particular, using the available experimental diphoton spectra, together with our SM and $B$-boson amplitudes, we determine which regions of the $\alpha_{B}$-$m_{B}$ plane are preferred by the data and assess the $B$-boson contribution.
It can be anticipated that, whilst the constraint from the $\eta\to\pi^0\gamma\gamma$ process in the resonant mass range $m_{\pi^{0}}\lesssim m_{B}\lesssim m_{\eta}$ is so strong that it makes it very difficult to identify any $B$-boson signatures
(assuming that the $B$ boson only decays into SM particles),
its imprint in the $t$ and $u$ channels may be noticeable in the invariant mass distributions when $m_{B}\lesssim m_{\pi^{0}}$ and $m_{B}\gtrsim m_{\eta}$, as it occurs with the nonresonant SM $\rho$, $\omega$ and $\phi$ exchanges~\cite{Escribano:2018cwg}.
Searches for leptophobic $B$ bosons require experimental precision, in order to disentangle their contribution from the SM, but also robust theoretical predictions. We 
attempt to undertake the latter in this work. 

This article is structured as follows.
In Sec.~\ref{subsection21}, we summarize the vector and scalar meson resonance exchange contributions to the amplitude for the three $\eta^{(\prime)}\to\pi^{0}\gamma\gamma$ and $\eta^{\prime}\to\eta\gamma\gamma$ decays \cite{Escribano:2018cwg}. In Sec.~\ref{subsection22}, we present the framework to include the contribution of intermediate $B$-boson exchanges to the amplitude.
We then use the above amplitudes in Sec.~\ref{section3} to, first, set limits on the $B$-boson parameters $\alpha_{B}$ and $m_{B}$ from the experimental branching ratios, and, second, to study the $B$-boson effect on the $\gamma\gamma$ and $\pi^{0}\gamma$ invariant mass spectra.
We conclude this work with some conclusions in Sec.~\ref{sectionConclusions}.


\section{Theoretical Framework}\label{section2}

\subsection{Standard Model: Vector and scalar contributions}\label{subsection21}

VMD and the L$\sigma$M can be used to calculate the SM contributions from vector and scalar meson resonance exchanges to the $\eta^{(\prime)}\to\pi^{0}\gamma\gamma$ and $\eta^{\prime}\to\eta\gamma\gamma$ decay processes.
In Ref.~\cite{Ametller:1991dp}, it was found that the VMD amplitude represents the dominant contribution to the $\eta\to\pi^0\gamma\gamma$ decay, whilst in~\cite{Escribano:2018cwg} we showed that this is also the case for the $\eta^\prime\to\pi^0\gamma\gamma$ and $\eta^\prime\to\eta\gamma\gamma$ processes.

In the VMD picture, the decay $\eta\to\pi^{0}\gamma\gamma$ proceeds through the transition $\eta\to V\gamma$ followed by $V\to\pi^{0}\gamma$, resulting in a total of six diagrams contributing to the amplitude of the process, which corresponds to the exchange of the three neutral vector mesons $V=\rho^{0},\omega$ and $\phi$ in the $t$ and $u$ channels. 
By combining the $V\eta\gamma$ and $V\pi^{0}\gamma$ interacting terms 
with the propagator of the exchanged vector mesons, one can calculate the vector meson contributions to the $\eta\to\pi^{0}\gamma\gamma$ decay.
We found~\cite{Escribano:2018cwg}
\begin{widetext}
\begin{eqnarray}
\label{AVMDetapi0}
\quad {\cal A}^{\mathrm{VMD}}_{\eta\to\pi^0\gamma\gamma}=
\sum_{V=\rho^0, \omega, \phi}g_{V\!\eta\gamma}g_{V\!\pi^0\gamma}
\left[\frac{(P\cdot q_2-m_\eta^2)\{a\}-\{b\}}{D_V(t)}
+\left\{
\begin{array}{c}
q_2\leftrightarrow q_1\\
t\leftrightarrow u
\end{array}
\right\}\right]\ ,
\end{eqnarray}
\end{widetext}
where
$t,u=(P-q_{2,1})^2=m_\eta^2-2P\cdot q_{2,1}$ are Mandelstam variables,
$\{a\}$ and $\{b\}$ are the Lorentz structures defined as
\begin{equation}
\label{LorentzStruct}
\begin{aligned}
\{a\}&=(\epsilon_1\cdot\epsilon_2)(q_1\cdot q_2)-(\epsilon_1\cdot q_2)(\epsilon_2\cdot q_1)\ ,\\[1ex]
\{b\}&=(\epsilon_1\cdot q_2)(\epsilon_2\cdot P)(P\cdot q_1)+(\epsilon_2\cdot q_1)
(\epsilon_1\cdot P)(P\cdot q_2)\\
&-(\epsilon_1\cdot\epsilon_2)(P\cdot q_1)(P\cdot q_2)-(\epsilon_1\cdot P)(\epsilon_2\cdot P)
(q_1\cdot q_2)\ ,
\end{aligned}
\end{equation}
where $P$ is the four-momentum of the decaying $\eta$ meson, and $\epsilon_{1,2}$ and $q_{1,2}$ are the polarization and four-momentum vectors of the final photons, respectively. 
The denominator $D_V(q^2)=m_V^2-q^2-i\,m_V\Gamma_V$ is the vector meson propagator, with $V=\rho^0$, $\omega$ and $\phi$. Due to the fact the the $\rho^0$ meson has got a very large decay width, the use of the usual Breit-Wigner prescription is not justified and, thus, one is compelled to make use of an energy-dependent decay width 
\begin{equation}
\label{varGamma}
\Gamma_{\rho^0}(q^2)=\Gamma_{\rho^0}
\left(\frac{q^2-4m_\pi^2}{m_{\rho^0}^2-4m_\pi^2}\right)^{3/2}\theta(q^2-4m_\pi^2)\ .
\end{equation}
The amplitudes for the decays $\eta^\prime\to\pi^0\gamma\gamma$ and $\eta^\prime\to\eta\gamma\gamma$ have a similar structure to that of Eq.~(\ref{AVMDetapi0}) with the replacements $m_\eta^2\to m_{\eta^\prime}^2$, and $g_{V\eta\gamma}g_{V\pi^0\gamma}\to g_{V\eta^\prime\gamma}g_{V\pi^0\gamma}$ for the $\eta^\prime\to\pi^0\gamma\gamma$ case and $g_{V\eta\gamma}g_{V\pi^0\gamma}\to g_{V\eta^\prime\gamma}g_{V\eta\gamma}$ for the $\eta^\prime\to\eta\gamma\gamma$ one.

For our analysis, we fix the $g_{VP\gamma}$ couplings in Eq.~(\ref{AVMDetapi0}) 
from experiment as follows: 
we first calculate the decay widths for the radiative transitions $V\to P\gamma$ and $P\to V\gamma$, 
and find the following relationships
\begin{equation}
\begin{aligned}
\Gamma_{V\to P\gamma}&=
\frac{1}{3}\frac{g^{2}_{VP\gamma}}{32\pi}\left(\frac{m_{V}^{2}-m_{P}^{2}}{m_{V}}\right)^{3}\ ,\\
\Gamma_{P\to V\gamma}&=
\frac{g^{2}_{VP\gamma}}{32\pi}\left(\frac{m_{P}^{2}-m_{V}^{2}}{m_{P}}\right)^{3}\ ,
\end{aligned}
\label{empirircalcoupling}
\end{equation}
which then are used 
in combination with the experimental decay widths from the PDG~\cite{Zyla:2020zbs}
to obtain the empirical $g_{VP\gamma}$ couplings provided in Table~\ref{gVPgammacouplingsempirical}.

\begin{table}[b]
\centering
\begin{tabular}{llllll}
\hline
Decay &\qquad BR &\qquad $|g_{V\!P\gamma}|$ GeV$^{-1}$\\
\hline
$\rho^0\to\pi^{0}\gamma$			&\qquad $(4.7\pm0.8)\times10^{-4}$		&\qquad $0.22(2)$\\ 
$\rho^0\to\eta\gamma$				&\qquad $(3.00\pm0.21)\times10^{-4}$	&\qquad $0.48(2)$\\ 
$\eta^{\prime}\to\rho^0\gamma$	&\qquad $(29.5\pm0.5)\%$			&\qquad $0.39(1)$\\ 
$\omega\to\pi^{0}\gamma$		&\qquad $(8.34\pm0.26)\%$			&\qquad $0.71(1)$\\ 
$\omega\to\eta\gamma$			&\qquad $(4.5\pm0.4)\times10^{-4}$		&\qquad $0.136(6)$\\ 
$\eta^{\prime}\to\omega\gamma$	&\qquad $(2.52\pm0.07)\%$			&\qquad $0.122(2)$\\ 
$\phi\to\pi^{0}\gamma$			&\qquad $(1.32\pm0.06)\times10^{-3}$	&\qquad $0.041(1)$\\ 
$\phi\to\eta\gamma$				&\qquad $(1.303\pm0.025)\%$			&\qquad $0.2093(20)$\\ 
$\phi\to\eta^{\prime}\gamma$		&\qquad $(6.22\pm0.21)\times10^{-5}$	&\qquad $0.216(4)$\\ 
\hline     
\end{tabular}
\caption{PDG values \cite{Zyla:2020zbs} for the branching ratios of the $V(P)\to P(V)\gamma$ transitions and the calculated $g_{V\!P\gamma}$ couplings directly from experiment (see Eq.~(\ref{empirircalcoupling}) and associated text).}
\label{gVPgammacouplingsempirical}
\end{table}

It is important to note that the most general $VP\gamma$ couplings in Eq.~(\ref{AVMDetapi0}) 
are energy dependent, i.e.~$g_{VP\gamma}(q^2)$. 
In the conventional VMD model, 
pseudoscalar mesons do not couple directly to photons but through the exchange of intermediate vectors;
thus, in this framework, a particular $VP\gamma$ coupling constant
times its normalized form factor 
is given by\footnote{Should $q^2$ be timelike, that is, $q^2>0$, 
then an imaginary part would need to be added to the propagator; 
this introduces the associated resonance width effects and rids the propagator from
its divergent behavior.}
\begin{equation}
\label{VMDcouplings}
g_{VP\gamma}\,\hat F_{VP\gamma}(q^2)=
\sum_{V^\prime}\frac{g_{VV^\prime P}\,g_{V^\prime\gamma}}{m_{V^\prime}^2-q^2}\ ,
\end{equation}
where $g_{VV^\prime P}$ are the vector-vector-pseudoscalar couplings,
$g_{V^\prime\gamma}$ the vector-photon conversion couplings,
and $m_{V^\prime}$ the intermediate vector masses.
In the $SU(3)$-flavor symmetry and OZI-rule respecting limits, 
one could express all the $g_{VP\gamma}$ in terms of
a single coupling constant $g$ and $SU(3)$-group factors~\cite{Bramon:1994pq}.
On the other hand, in the context of resonance chiral theory (RChT), for instance, the $VP\gamma$ effective vertex is made of a local $VP\gamma$ vertex weighted by a coupling constant, $h_V$, and a nonlocal one built from the exchange of an intermediate vector weighted by a second coupling constant, $\sigma_V$, times the vector-photon conversion factor $f_V$~\cite{Prades:1993ys}
\begin{equation}
\label{RChTcouplings}
\begin{aligned}
g_{VP\gamma}\,\hat F_{VP\gamma}(q^2)&=
C_{VP\gamma}\lvert e\rvert \frac{4\sqrt{2}\,h_V}{f_\pi}\\
&\quad\times\left(1+\frac{\sigma_V f_V}{\sqrt{2}\,h_V}
\frac{q^2}{m_{V^\prime}^2-q^2}\right)\ ,
\end{aligned}
\end{equation}
where $C_{VP\gamma}$ are $SU(3)$-group factors.
Notwithstanding this, and down to the fact that the outgoing photons in 
$\eta^{(\prime)}\to \pi^0\gamma\gamma$ and $\eta^{\prime}\to \eta\gamma\gamma$ are on shell, 
the energy dependence of the vertex form factors vanish in either model and the corresponding 
couplings become just constants~\cite{Escribano:2020rfs}. 
As has just been explained, 
in the approach followed in the present work we extract the values for these coupling constants 
directly from experiment not relying on any specific model~\cite{Bramon:1994pq,Prades:1993ys}, 
rendering our theoretical treatment of vector exchanges rather model independent. 

Moving on to the effects of scalar resonance exchanges on the decays under study, they were explicitly assessed in Ref.~\cite{Escribano:2018cwg} making use of the L$\sigma$M and it was found that the scalar contributions are subdominant, whilst the exchange of vector resonances largely dominate. Accordingly, it has been deemed not necessary to employ a more sophisticated theoretical treatment, such as dispersive methods (see e.g.~Refs.~\cite{Oset:2002sh,Oset:2008hp}), to analyze the effect of the scalar exchanges in this work. This, of course, would not be possible for processes where the contribution from scalar resonances is dominant, for instance in $\phi \to \eta \pi^0 \gamma$~\cite{Escribano:2006mb,Moussallam:2021dpk}, and an improved theoretical treatment would, therefore, be required in those cases.

\subsection{Beyond the Standard Model: $B$-boson contribution}\label{subsection22}

In analogy to the VMD contributions summarized in the previous subsection, 
we next define the framework to include intermediate $B$-boson exchanges to the decay amplitude. 

\begin{figure}[h]
\centering\includegraphics[scale=0.55]{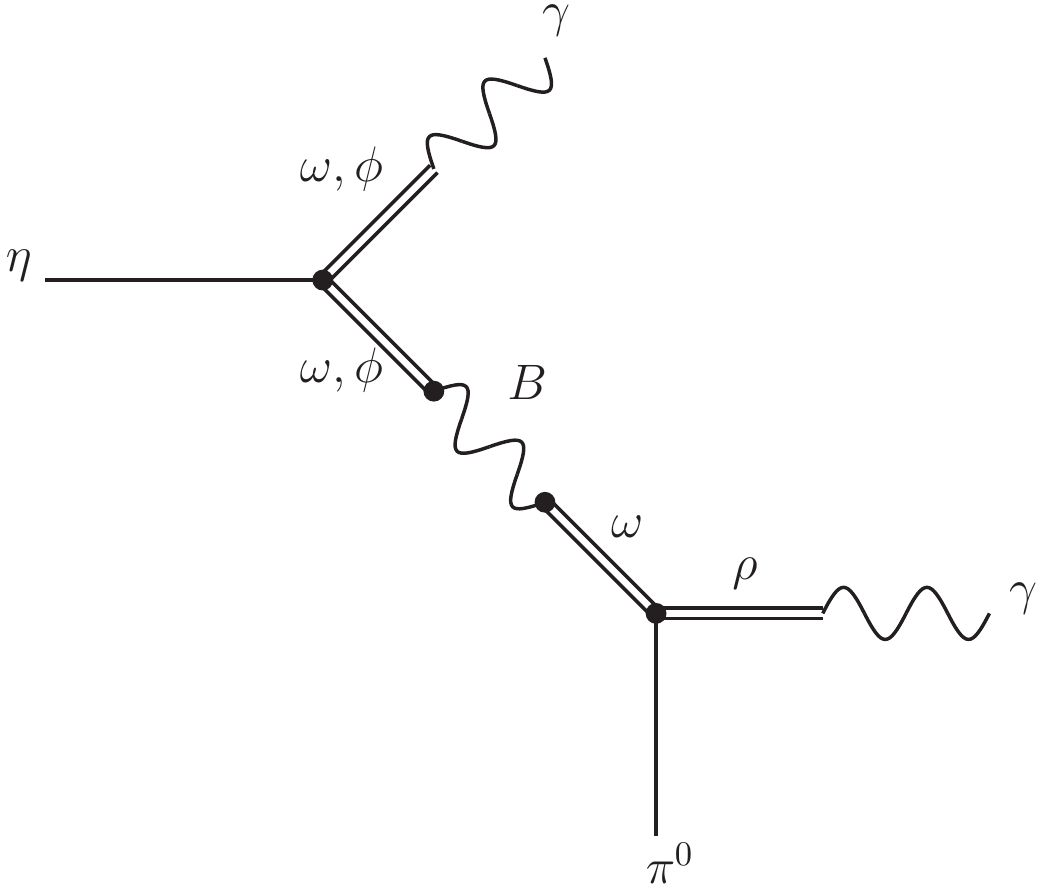}
\caption{Schematic diagram of the $B$-boson exchange mechanism for the decay $\eta\to\pi^{0}\gamma\gamma$.}
\label{Fig:BbosonExchange} 
\end{figure}

The diagrammatic representation of the decay process is depicted in Fig.~\ref{Fig:BbosonExchange} for the $\eta\to\pi^{0}\gamma\gamma$ case.\footnote{It should be mentioned that the same diagram where the $B$ boson is replaced by a photon also exists. However, this is not considered in the present analysis given that this contribution is highly suppressed with respect to the intermediate vector exchanges that has already been considered in Sec.~\ref{subsection21} and introduces unnecessary complexity.}
This contribution can be assessed from the conventional VMD $VVP$ and $V\gamma$ 
Lagrangians~\cite{Bramon:1992kr}
\begin{equation}
\begin{aligned}
\label{Eq:VPgammaLagrangian}
\mathcal{L}_{VVP}&=\frac{G}{\sqrt{2}}\epsilon^{\mu\nu\alpha\beta}
{\rm{tr}}\left[\partial_{\mu}V_{\nu}\partial_{\alpha}V_{\beta}P\right]\ ,\\[1ex]
\mathcal{L}_{V\gamma}&=-4egf_{\pi}^{2}A^{\mu}{\rm{tr}}\left[QV_{\mu}\right]\ ,
\end{aligned}
\end{equation} 
where $G=3g^{2}/4\pi^{2}f_{\pi}$, 
$V^{\mu}$ and $P$ are the matrices for the vector and pseudoscalar meson fields,
$A^{\mu}$ is the photon field, and $Q={\rm{diag}}\{2/3,-1/3,-1/3\}$ is the quark-charge matrix,
supplemented by an effective Lagrangian that describes the $VB$ interaction.
The latter is formally
identical to the $V\gamma$ Lagrangian 
with the substitutions $A^{\mu}\to B^{\mu}$, $e\to g_{B}$ 
and $Q\to{\rm{diag}}\{1/3,1/3,1/3\}$, and it is given by
\begin{eqnarray}
\label{Eq:Lagrangians3}
\mathcal{L}_{VB}=-4\frac{1}{3}g_{B}gf_{\pi}^{2}B^{\mu}{\rm{tr}}\left[V_{\mu}\right]\ .
\end{eqnarray}

From the $VVP$ and $VB$ Lagrangians 
in Eqs.~(\ref{Eq:VPgammaLagrangian}) and (\ref{Eq:Lagrangians3}), respectively, 
along with the corresponding $V$-meson propagators, 
it is straightforward to obtain expressions for the $g_{BP\gamma}$ couplings 
in terms of the generic $B$-boson coupling $g_{B}$.
The $g_{BP\gamma}$ couplings are energy dependent 
and read
\begin{equation}
\label{Eq:gBPgammaCouplings}
\begin{aligned}
g_{B\pi^{0}\gamma}(q^2)&=\frac{eg_{B}}{4\pi^{2}f_{\pi}}F_{\omega}(q^2)\ ,\\[1ex]
g_{B\eta\gamma}(q^2)&=\frac{eg_{B}}{12\pi^{2}f_{\pi}}
\left[{\rm c}\varphi_{P}F_{\omega}(q^2)+\sqrt{2}{\rm s}\varphi_{P}F_{\phi}(q^2)\right]\ ,\\[1ex]
g_{B\eta^{\prime}\gamma}(q^2)&=\frac{eg_{B}}{12\pi^{2}f_{\pi}}
\left[{\rm s}\varphi_{P} F_{\omega}(q^2)-\sqrt{2}{\rm c}\varphi_{P}F_{\phi}(q^2)\right]\ ,
\end{aligned}
\end{equation}
where $\varphi_{P}$ is the $\eta$-$\eta^{\prime}$ mixing angle 
in the quark-flavor basis~\cite{Bramon:1997va}
and the abbreviations $c\varphi_{P}\equiv\cos\varphi_{P}$ and $s\varphi_{P}\equiv\sin\varphi_{P}$
have been employed.
The functions $F_{V}(q^2)$ in the previous equations are form factors that account for the 
$\omega$ and $\phi$ propagation, and are given by
\begin{equation}
F_{V}(q^2)=\frac{m_{V}^{2}}{m_{V}^{2}-q^2-im_{V}\Gamma_{V}}\ .    
\end{equation}

Combining the $g_{B\pi^{0}\gamma}$ and $g_{B\eta\gamma}$ couplings 
from Eq.~(\ref{Eq:gBPgammaCouplings}) with the propagator of the $B$ boson, 
allows one to find the $B$-boson exchange contribution to the amplitude of the
$\eta\to\pi^{0}\gamma\gamma$ decay
\begin{widetext}
\begin{eqnarray}
\label{Eq:BbosonAmplitude}
{\cal A}^{B\,\mathrm{boson}}_{\eta\to\pi^0\gamma\gamma}=
g_{B\eta\gamma}(t)g_{B\pi^0\gamma}(t)
\left[\frac{(P\cdot q_2-m_\eta^2)\{a\}-\{b\}}{D_B(t)}+
\bigg\{
\begin{array}{c}
q_2\leftrightarrow q_1\\
t\leftrightarrow u
\end{array}
\bigg\}\right]\ ,
\end{eqnarray}
\end{widetext}
where $D_B(q^2)=m_{B}^{2}-q^2-im_{B}\Gamma_{B}$ is the $B$-boson propagator.
The $B$-boson contribution to the amplitudes of the $\eta^\prime\to\pi^0\gamma\gamma$ and $\eta^\prime\to\eta\gamma\gamma$ decays have a similar structure to that of Eq.~(\ref{Eq:BbosonAmplitude}) with the replacements $m_\eta^2\to m_{\eta^\prime}^2$, and $g_{B\eta\gamma}g_{B\pi^0\gamma}\to g_{B\eta^\prime\gamma}g_{B\pi^0\gamma}$ for the $\eta^\prime\to\pi^0\gamma\gamma$ and $g_{B\eta\gamma}g_{B\pi^0\gamma}\to g_{B\eta^\prime\gamma}g_{B\eta\gamma}$ for the $\eta^\prime\to\eta\gamma\gamma$.

The decay widths for the radiative transitions $\eta^{(\prime)}\to B\gamma$ and $B\to\pi^{0}\gamma,\eta^{(\prime)}\gamma$ can be calculated from 
Eq.~(\ref{Eq:gBPgammaCouplings}) and the analogous to Eq.~(\ref{empirircalcoupling}).
They are given by
\begin{equation}
\label{Eq:DecayWidthPtoBgamma}
\begin{aligned}
\Gamma_{\eta\to B\gamma}&=\frac{\alpha_{em}\alpha_{B}m_{\eta}^{3}}
{288\pi^{3}f_{\pi}^{2}}\left(1-\frac{m_{B}^{2}}{m_{\eta}^{2}}\right)^{3}\\
&\quad\times\left[{\rm c}\varphi_{P}F_{\omega}(m_{B}^{2})+
\sqrt{2}{\rm s}\varphi_{P}F_{\phi}(m_{B}^{2})\right]^{2}\ ,\\[1ex]
\Gamma_{\eta^{\prime}\to B\gamma}=&\frac{\alpha_{em}\alpha_{B}m_{\eta^{\prime}}^{3}}
{288\pi^{3}f_{\pi}^{2}}\left(1-\frac{m_{B}^{2}}{m_{\eta^{\prime}}^{2}}\right)^{3}\\
&\quad\times\left[{\rm s}\varphi_{P}F_{\omega}(m_{B}^{2})
-\sqrt{2}{\rm c}\varphi_{P}F_{\phi}(m_{B}^{2})\right]^{2}\ ,
\end{aligned}
\end{equation}
for the $B$ production from $\eta^{(\prime)}$ decays and 
\begin{equation}
\label{Eq:DecayWidthBtoPgamma}
\begin{aligned}
\Gamma_{B\to\pi^{0}\gamma}&=\frac{\alpha_{em}\alpha_{B}m_{B}^{3}}{96\pi^{3}f_{\pi}^{2}}
\left(1-\frac{m_{\pi}^{2}}{m_{B}^{2}}\right)^{3}|F_{\omega}(m_{B}^{2})|^{2}\ ,\\[1ex]  
\Gamma_{B\to\eta\gamma}&=\frac{\alpha_{em}\alpha_{B}m_{B}^{3}}
{864\pi^{3}f_{\pi}^{2}}\left(1-\frac{m_{\eta}^{2}}{m_{B}^{2}}\right)^{3}\\
&\quad\times\left[{\rm c}\varphi_{P}F_{\omega}(m_{B}^{2})
+\sqrt{2}{\rm s}\varphi_{P}F_{\phi}(m_{B}^{2})\right]^{2}\ ,\\[1ex]
\Gamma_{B\to\eta^{\prime}\gamma}&=\frac{\alpha_{em}\alpha_{B}m_{B}^{3}}
{864\pi^{3}f_{\pi}^{2}}\left(1-\frac{m_{\eta^{\prime}}^{2}}{m_{B}^{2}}\right)^{3}\\
&\quad\times\left[{\rm s}\varphi_{P}F_{\omega}(m_{B}^{2})
-\sqrt{2}{\rm c}\varphi_{P}F_{\phi}(m_{B}^{2})\right]^{2}\ ,
\end{aligned}
\end{equation}
for the $B$-boson decays.
The leptonic decays arise from the kinetic mixing of the $B$ boson with the photon, 
cf.~Eq.~(\ref{Eq:BbosonLagrangian}), and read~\cite{Tulin:2014tya}
\begin{equation}
\Gamma_{B\to\ell^{+}\ell^{-}}=\frac{\alpha_{em}\varepsilon^{2}m_{B}}{3}
\left(1+\frac{2m_{\ell}^{2}}{m_{B}^{2}}\right)\sqrt{1-\frac{4m_{\ell}^{2}}{m_{B}^{2}}}\ ,
\label{Eq:DecayWidthBtoLeptons}
\end{equation}
whilst the $B$-boson decay to $\pi^{+}\pi^{-}$, which also depends on $\epsilon$, 
is given by~\cite{Tulin:2014tya}
\begin{equation}
\Gamma_{B\to\pi^{+}\pi^{-}}=\frac{\alpha_{em}\varepsilon^{2}m_{B}}{12}
\left(1-\frac{4m_{\pi}^{2}}{m_{B}^{2}}\right)^{3/2}|F_{\pi}(m_{B}^{2})|^{2}\ ,
\label{Eq:DecayWidthBto2pions}
\end{equation}
where $F_{\pi}(q^2)$ is the pion vector form factor.
Finally, for the three-body decay $\Gamma_{B\to\pi^{+}\pi^{-}\pi^{0}}$
we make use of the following amplitude
\begin{widetext}
\begin{equation}
\quad{\cal A}^{B\,\mathrm{boson}}_{B\to\pi^+\pi^-\pi^0}=
\frac{g^2g_B}{\pi^2 f_{\pi}}\epsilon_{\mu\nu\alpha\beta}
\left(\frac{p_+^\mu p_-^\nu p_0^\alpha}{D_{\rho^0}(s)}
+\frac{p_-^\mu p_0^\nu p_+^\alpha}{D_{\rho^-}(t)}
+\frac{p_0^\mu p_+^\nu p_-^\alpha}{D_{\rho^+}(u)}\right)\varepsilon^{\beta}F_{\omega}(m_{B}^{2})\ ,
\label{ABpipipi}
\end{equation}
\end{widetext}
where $\varepsilon_{\beta}$ is the polarization vector of the $B$ boson, 
$D_{\rho}(q^2)$ is the $\rho$ propagator with energy-dependent width defined in Eq.~(\ref{varGamma}),
$p_+$, $p_-$ and $p_0$ are the four-momentum vectors associated to the $\pi^+$, $\pi^-$ and $\pi^0$, 
respectively, and the Mandelstam variables $s$, $t$ and $u$ are defined, in this instance, as 
$s=(p_++p_-)^2$, $t=(p_-+p_0)^2$ and $u=(p_++p_0)^2$.


\section{Limits on $\alpha_{B}$ and $m_{B}$}\label{section3}

In this section, we make use of the theoretical expressions presented in Secs.~II A and II B, along with the available experimental data, to place limits on the $B$-boson parameters $\alpha_B$ and $m_B$.

\begin{figure}
\centering\includegraphics[scale=0.375]{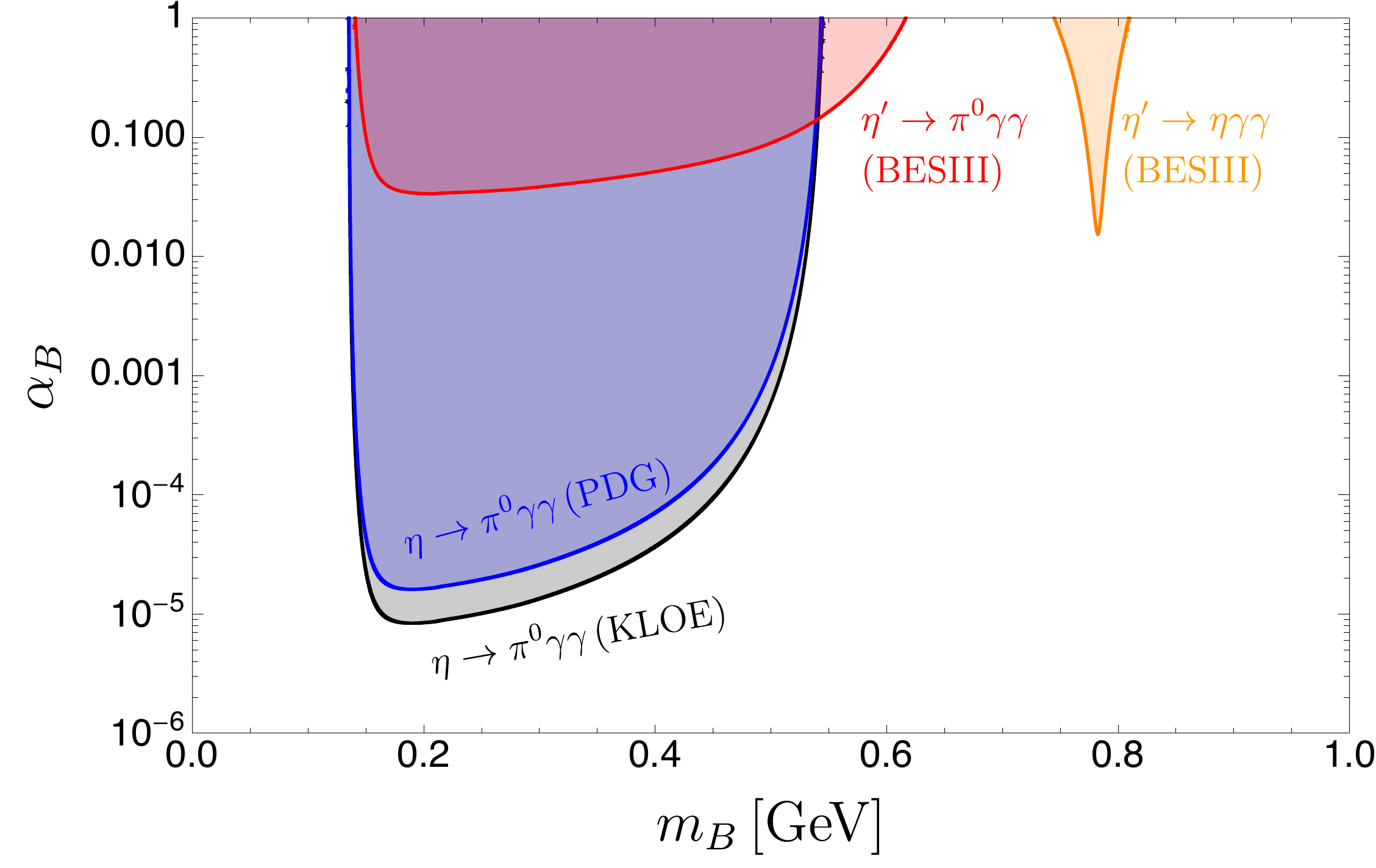}
\caption{Limits on the leptophobic $U(1)_{B}$-boson parameters $\alpha_{B}$ and $m_{B}$  
from the $\eta\to\pi^{0}\gamma\gamma$ BR measured by KLOE~\cite{Cao:2022rxo} (black line) 
and the value reported by the PDG~\cite{Zyla:2020zbs} (blue line). 
Also shown are the limits from the BESIII measurements of $\eta^{\prime}\to\pi^{0}\gamma\gamma$
(red line)~\cite{BESIII:2016oet} and $\eta^{\prime}\to\eta\gamma\gamma$ 
(orange line)~\cite{BESIII:2019ofm}. 
Following the approach of Ref.~\cite{Tulin:2014tya}, 
the SM contribution is set to zero in all cases and the NWA is applied.
The shaded regions are excluded.}
\label{Fig:Constraints} 
\end{figure}

As a preliminary step, we adopt the approach presented in Ref.~\cite{Tulin:2014tya} 
with the most up-to-date experimental data to generate 
limits on the $B$-boson parameters from the decays under study, 
$\eta^{(\prime)}\to\pi^{0}\gamma\gamma$ and $\eta^{\prime}\to\eta\gamma\gamma$, 
which are shown in the form of exclusion plots in Fig.~\ref{Fig:Constraints}.
That approach neglects the SM contribution and uses the NWA 
to place limits upon requiring that the $B$-boson contribution does not exceed the 
total observed branching ratio (BR) at $2\sigma$.
The curves for the $\eta\to\pi^{0}\gamma\gamma$ process 
come from the (preliminary) value found by the KLOE Collaboration, $\rm{BR}=(1.23\pm0.14)\times10^{-4}$~\cite{Cao:2022rxo} (black line), and the BR reported by the PDG, $\rm{BR}=(2.56\pm0.22)\times10^{-4}$~\cite{Zyla:2020zbs} (blue line); we also show the traces obtained from the BESIII Collaboration measurements for the decays $\eta^{\prime}\to\pi^{0}\gamma\gamma$, $\rm{BR}=(3.20\pm0.07\pm0.23)\times10^{-3}$~\cite{BESIII:2016oet} (red line), and $\eta^{\prime}\to\eta\gamma\gamma$, $\rm{BR}=(8.25\pm3.41\pm0.72)\times10^{-5}$~\cite{BESIII:2019ofm} (orange line).

\begin{figure}
\centering\includegraphics[scale=0.375]{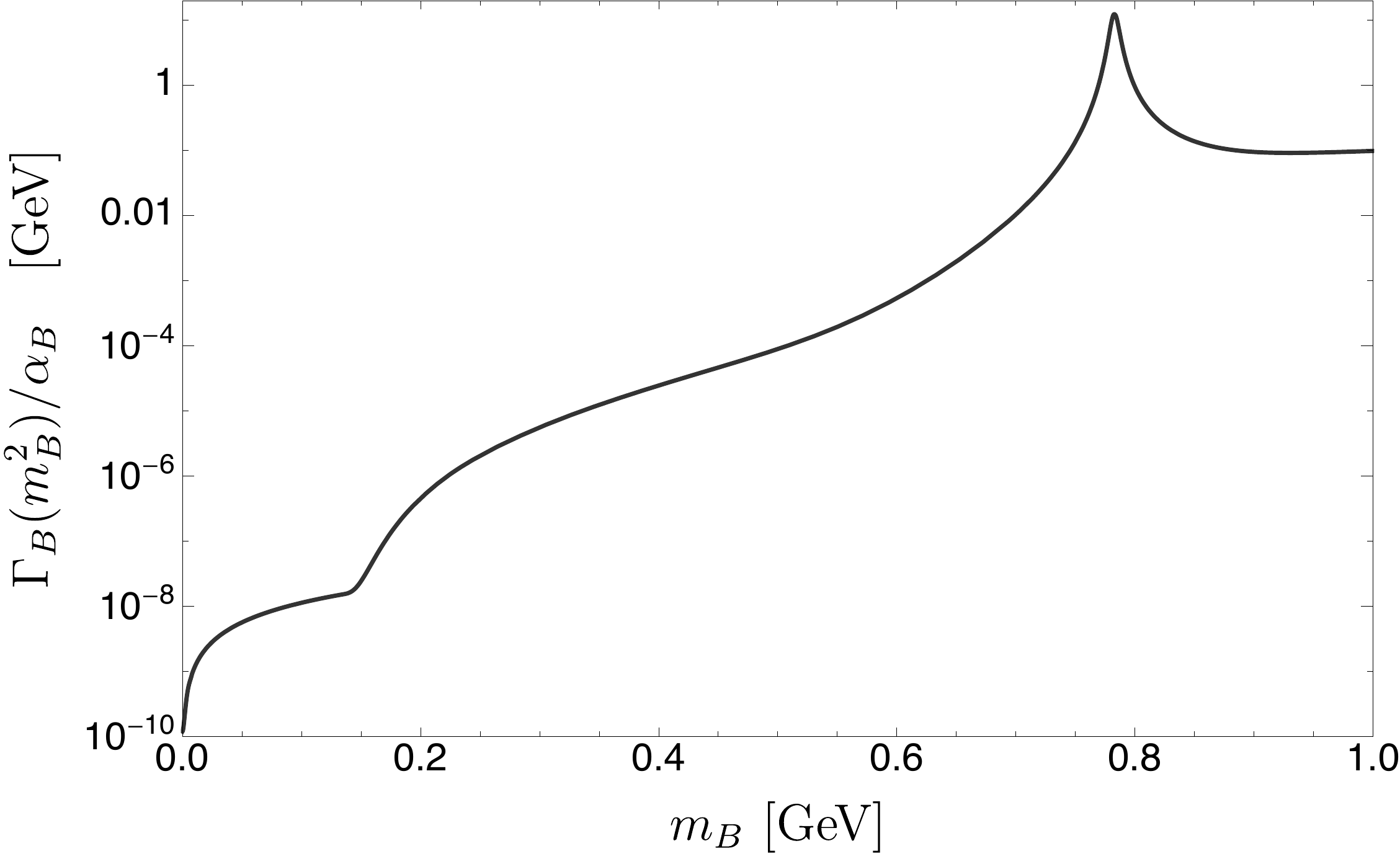}
\caption{Normalized width of the $B$ boson, $\Gamma_{B}(m_{B}^{2})/\alpha_B$, as a function of $m_{B}$ 
from Eq.~(\ref{Eq:DecayWidth2}).}
\label{Fig:BbosonTotalWidth} 
\end{figure}

The above limits can clearly be improved by including the contribution from the SM 
in the theoretical treatment. 
Taking this into account, the amplitude for these decay processes is written as the coherent sum 
of the vector, scalar and $B$-boson exchange contributions, 
$\mathcal{A}=\mathcal{A}_{\rm{VMD}}+\mathcal{A}_{\rm{L\sigma M}}+\mathcal{A}_{B\,\rm{ boson}}$ 
(cf.~Secs.~\ref{subsection21}, \ref{subsection22} and Ref.~\cite{Escribano:2018cwg}).
The corresponding partial decay widths 
depend on a total of three parameters: 
i) the baryonic fine-structure constant, $\alpha_{B}$, 
ii) the $B$-boson mass, $m_{B}$, and 
iii) its total decay width, $\Gamma_{B}$.
However, given that $\Gamma_{B}$ is not an independent parameter 
(that is, it can be expressed in terms of $\alpha_{B}$ and $m_{B}$), 
we can reduce the number of free parameters from three to two. 
Accordingly, the denominator in Eq.~(\ref{Eq:BbosonAmplitude}), $D_B(q^2)$, is replaced by
\begin{equation}
\mathcal{D}_{B}(q^2)=m_{B}^{2}-q^2-i\sqrt{q^2}\,\Gamma_{B}(q^2)\ ,
\label{Eq:BWBboson}
\end{equation}
where $\Gamma_{B}(q^2)=\sum_{i}\Gamma_{B}^{i}(q^2)$ is the energy-dependent width of the $B$ boson, 
with the sum running over the partial widths of the various decay channels the $B$ boson can decay into.
For our study, we include the partial widths of the decay channels 
$B\to\pi^{0}\gamma$, $e^{+}e^{-}$, $\mu^{+}\mu^{-}$, and $\pi^{+}\pi^{-}$ given, respectively, 
in Eqs.~(\ref{Eq:DecayWidthBtoPgamma})--(\ref{Eq:DecayWidthBto2pions}),
whilst for the partial width of the $B\to\pi^{0}\pi^{+}\pi^{-}$ 
we make use of the amplitude in Eq.~(\ref{ABpipipi}) to obtain numerical results 
after squaring and numerically integrating over its corresponding phase space. 
The energy-dependent width $\Gamma_B(q^2)$ can, therefore, be written as
\begin{widetext}
\begin{equation}
\begin{aligned}
\Gamma_{B}(q^2)&=\theta(q^2-m_{\pi}^{2})
\frac{{\gamma}_{B\to\pi^{0}\gamma}(q^2)}{{\gamma}_{B\to\pi^{0}\gamma}(m_{B}^{2})}
\Gamma_{B\to\pi^{0}\gamma}
+\theta(q^2-4m_{e}^{2})
\frac{{\gamma}_{B\to e^{+}e^{-}}(q^2)}{{\gamma}_{B\to e^{+}e^{-}}(m_{B}^{2})}
\Gamma_{B\to e^{+}e^{-}}
+\theta(q^2-4m_{\mu}^{2})
\frac{{\gamma}_{B\to\mu^{+}\mu^{-}}(q^2)}{{\gamma}_{B\to\mu^{+}\mu^{-}}(m_{B}^{2})}
\Gamma_{B\to \mu^{+}\mu^{-}}\\[1ex]
&+\theta(q^2-4m_{\pi}^{2})
\frac{{\gamma}_{B\to\pi^{+}\pi^{-}}(q^2)}{{\gamma}_{B\to\pi^{+}\pi^{-}}(m_{B}^{2})}
\Gamma_{B\to \pi^{+}\pi^{-}}
+\theta(q^2-9m_{\pi}^{2})\frac{{\gamma}_{B\to\pi^{+}\pi^{-}\pi^{0}}(q^2)}{{\gamma}_{B\to\pi^{+}\pi^{-}\pi^{0}}(m_{B}^{2})}
\Gamma_{B\to\pi^{+}\pi^{-}\pi^{0}}\ ,
\label{Eq:DecayWidth2}
\end{aligned}
\end{equation}
\end{widetext}
where the ${\gamma}_{i}(q^2)$ parameters are given by the following expressions
\begin{equation}
\begin{aligned}
{\gamma}_{B\to\pi^{0}\gamma}(q^2)&=
(q^2)^{3/2}\left(1-\frac{m_{\pi}^{2}}{q^2}\right)^{3}|F_{\omega}(q^2)|^{2}\ ,\\[1ex]
{\gamma}_{B\to\ell^{+}\ell^{-}}(q^2)&=
\sqrt{q^2}\left(1+\frac{2m_{\ell}^{2}}{q^2}\right)\sqrt{1-\frac{4m_{\ell}^{2}}{q^2}}\ ,\\[1ex]
{\gamma}_{B\to\pi^{+}\pi^{-}}(q^2)&=
\sqrt{q^2}\left(1-\frac{4m_{\pi}^{2}}{q^2}\right)^{3/2}|F_{\pi}(q^2)|^{2}\ ,
\end{aligned}
\end{equation} 
whilst ${\gamma}_{B\to\pi^{+}\pi^{-}\pi^{0}}(q^2)$ 
must be evaluated numerically.
In Fig.~\ref{Fig:BbosonTotalWidth}, the total normalized width $\Gamma_{B}(m_{B}^{2})/\alpha_B$ 
is plotted as a function of $m_{B}$. 

\begin{figure}
\centering\includegraphics[scale=0.5]{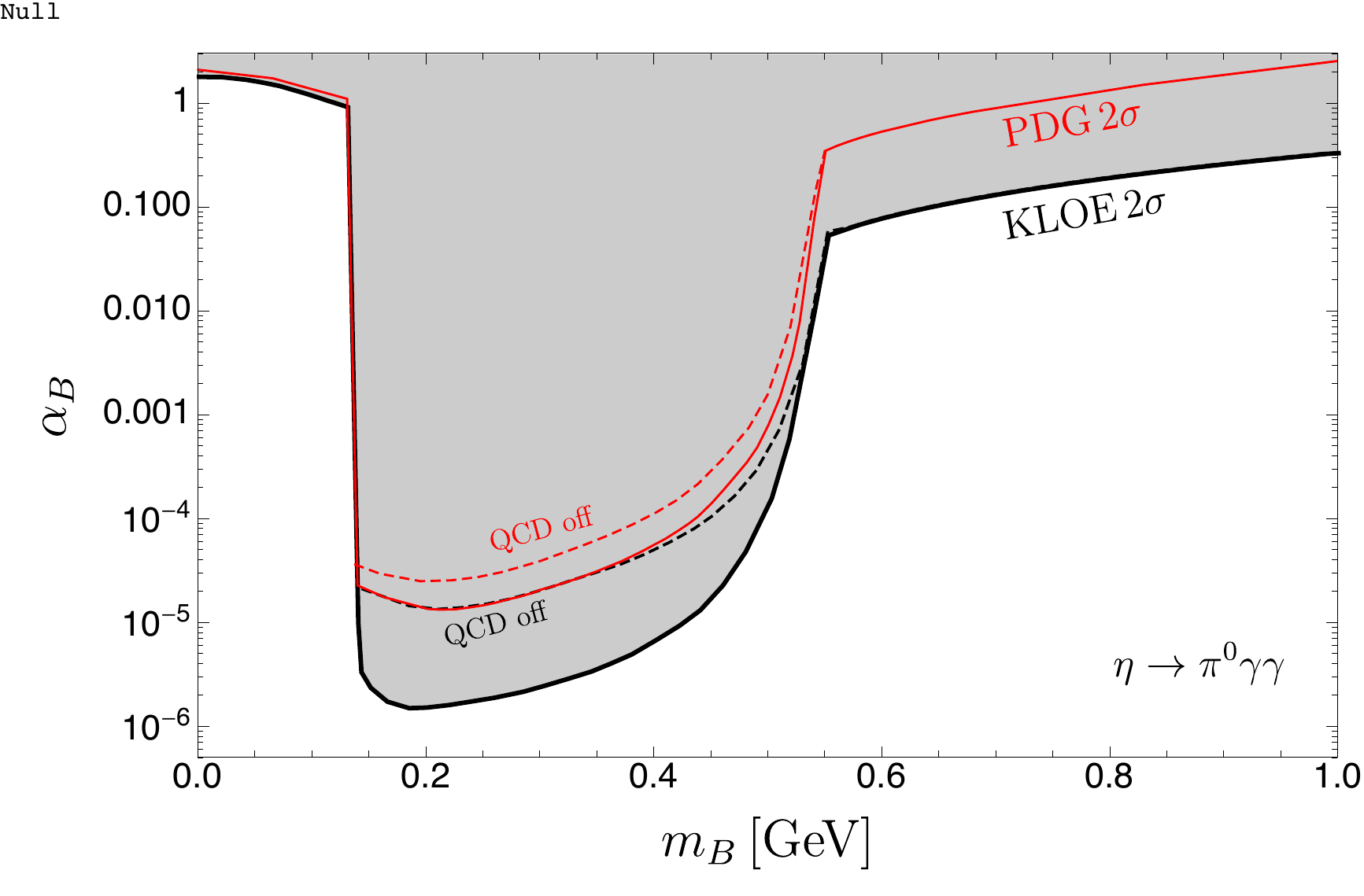}
\caption{Limits on the leptophobic $B$-boson coupling $\alpha_{B}$ for different $m_{B}$ masses from the $\eta\to\pi^{0}\gamma\gamma$ BR measurements by KLOE \cite{Cao:2022rxo} (black line) and the PDG \cite{Zyla:2020zbs} (red line).
The gray shaded region is excluded by KLOE and the dashed lines correspond to the limits with the QCD contributions turned off.}
\label{Fig:ExclusionPlot} 
\end{figure}

Next, we proceed to calculate the constraints on the $B$-boson parameters $\alpha_{B}$ and $m_{B}$ 
set by experiment.
We start with the $\eta\to\pi^{0}\gamma\gamma$ decay using the PDG reported value, 
$\rm{BR}=(2.56\pm0.22)\times10^{-4}$~\cite{Zyla:2020zbs},
as well as the (preliminary) value from the KLOE Collaboration, 
$\rm{BR}=(1.23\pm0.14)\times10^{-4}$~\cite{Cao:2022rxo} (see also Ref.~\cite{KLOECD2021}). 
In Fig.~\ref{Fig:ExclusionPlot}, we show the limits in the $\alpha_{B}$-$m_{B}$ plane, 
which are found by requiring our predictions 
to not exceed the corresponding branching ratios at $2\sigma$.
The gray area is excluded by the data from KLOE, 
which yield a more stringent limit than the resulting one from the PDG (solid red line).
This is as expected given that the BR from KLOE is found to be in good agreement with 
our SM prediction from Ref.~\cite{Escribano:2018cwg}, 
$\rm{BR}=(1.35\pm0.08)\times10^{-4}$, 
and, thus, the KLOE constraints on the $B$ boson turn out to be stronger.
The dashed black line in the figure is found using the data from KLOE 
but with the SM (or, equivalently, QCD) contributions set to zero.
Clearly, these contributions are not negligible as the limits on $\alpha_{B}$ 
become an order of magnitude weaker when their effects are turned off 
(labeled QCD off in the plots).  
The uncertainty in the exclusion limits associated to the systematic errors of our theoretical treatment is presented in Appendix~\ref{Appendix2}.

The shape and size of the excluded region in Fig.~\ref{Fig:ExclusionPlot} contains key physical information. 
In this figure, three different regions are observed. 
The first one corresponds to $m_{B}\lesssim m_{\pi^{0}}$, where $\alpha_{B}\sim\mathcal{O}(1)$.
At $m_{B}\sim m_{\pi^{0}}$, 
the limit placed on the coupling plummets by almost six orders of magnitude down to $\alpha_{B}\sim10^{-6}$; 
it then moderately increases, to finally take a steep rise when $m_{B}$ approaches $m_{\eta}$, 
reaching $\alpha_{B}\sim10^{-2}$.
Finally, for $m_{B}\gtrsim m_{\eta}$ the constraint on the coupling grows very smoothly as $m_B$ increases.
Out of the three, the $m_{\pi^{0}}\lesssim m_{B}\lesssim m_{\eta}$ region deserves special attention and 
raises the question as to why $\alpha_{B}$ is constrained so strongly there.
The answer to this is related to the fact that the $B$-boson width is extremely small in this region of parameter space.

Let us look into this in more detail. By noticing from Fig.~\ref{Fig:BbosonTotalWidth} that within the $m_{\pi^{0}}\lesssim m_{B}\lesssim m_{\eta}$ mass range the NWA is valid, it allows us to write the squared modulus of the $B$-boson propagator as $\pi/(m_{B}\Gamma_{B})\delta(t-m_{B}^{2})$, under the phase-space integral. For a $B$ boson whose squared mass falls within the kinematic space for the $t$ variable,  
i.e.~$t_{\rm{min}}\leq m_B^2\leq t_{\rm{max}}$, the phase-space integral over $dt$ places the $B$ boson on shell and one is allowed to write
\begin{equation}
\label{eq:etadecayprop}
\Gamma(\eta\to\pi^{0}\gamma\gamma)\!\propto\!\!\!\int\!\!\frac{\alpha_B^2\,dt}{|\mathcal{D}_{B}(t)|^2}\to\frac{\alpha_B^2\,\pi}{m_B\,\Gamma_B(m_B^2)}\ .
\end{equation} 
As it can be seen in Fig.~\ref{Fig:BbosonTotalWidth}, $\Gamma_B(m_B^2)/\alpha_B$ is very small within the kinematic region of interest for the present discussion (i.e.~$m_{\pi^{0}}\lesssim m_{B}\lesssim m_{\eta}$), which, in the $\Gamma_B(t)/\alpha_B\to0$ limit, forces $\alpha_B\to0$ so that $\Gamma(\eta\to\pi^{0}\gamma\gamma)$ remains finite.

\begin{figure*}
\includegraphics[scale=0.375]{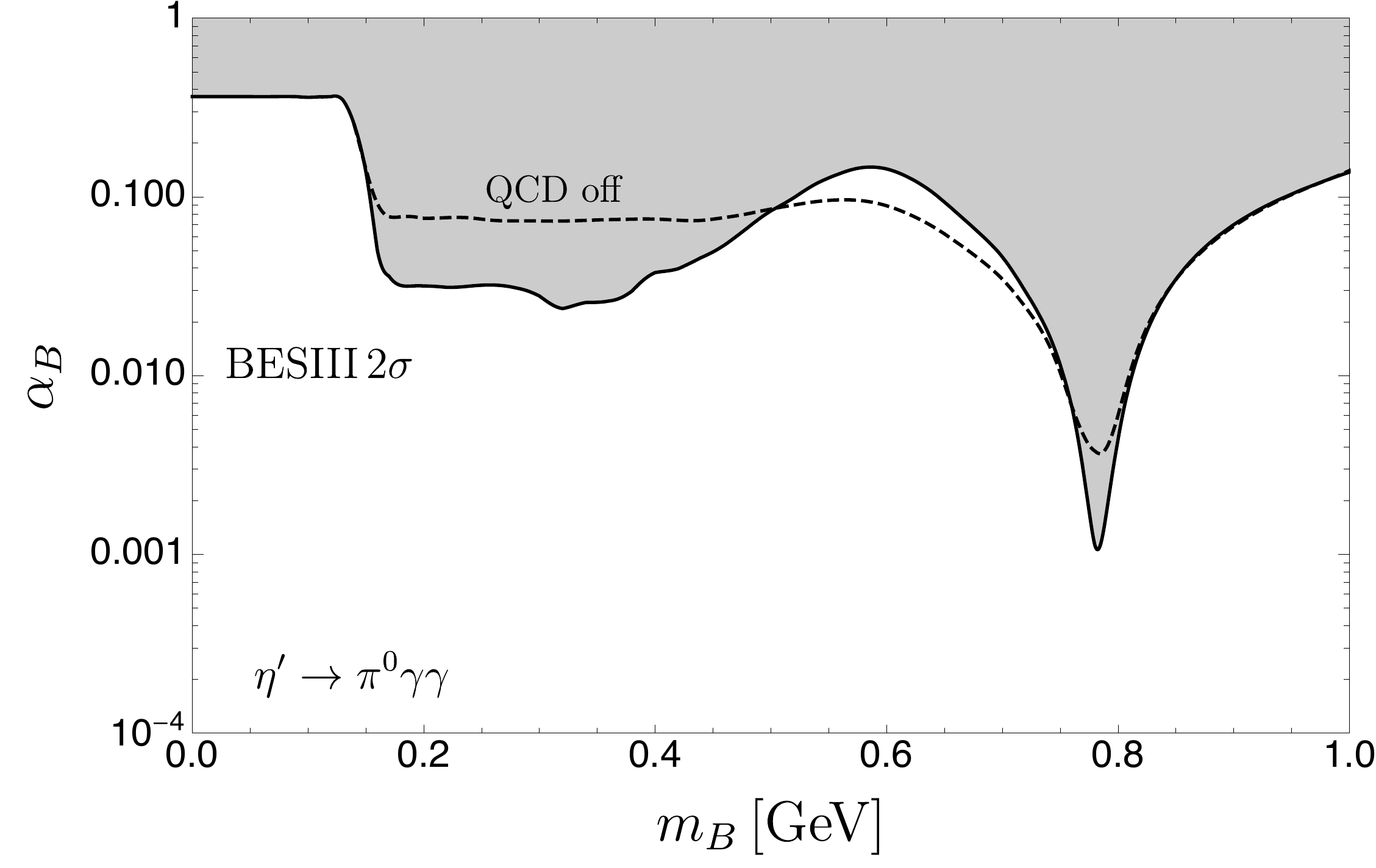}\,\qquad\includegraphics[scale=0.375]{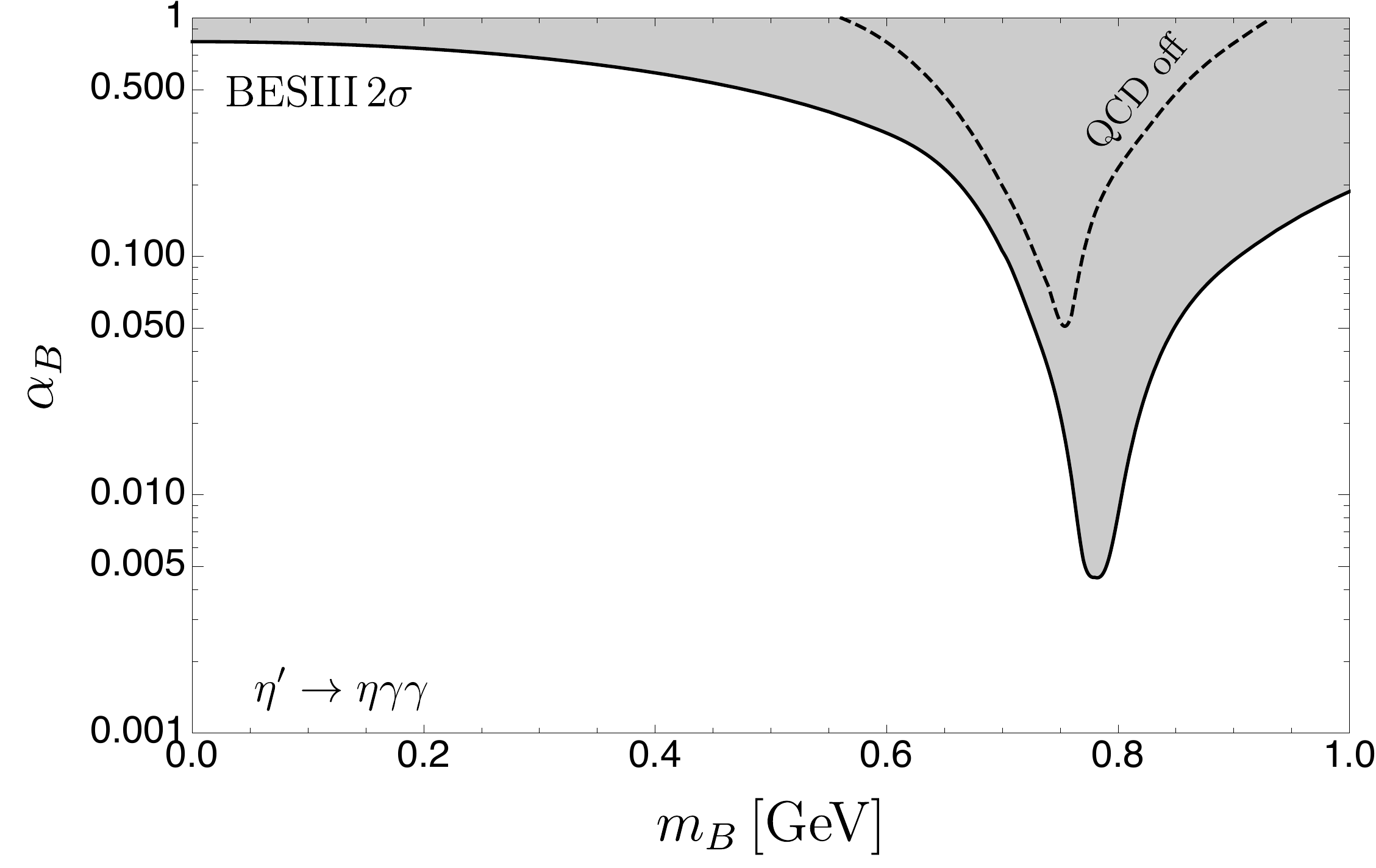}
\caption{Limits on the leptophobic $B$-boson coupling $\alpha_{B}$ for different $m_B$ masses from the BR measurements of the decays $\eta^{\prime}\to\pi^{0}\gamma\gamma$ (left plot)~\cite{BESIII:2016oet} and $\eta^{\prime}\to\eta\gamma\gamma$ (right plot)~\cite{BESIII:2019ofm} by BESIII.
The gray shaded region is excluded and the dashed black line corresponds to the limit with the QCD contributions set to zero.}
\label{Fig:ExclusionPlot2} 
\end{figure*}

Next, we show the exclusion plots associated to the two $\eta^{\prime}$ decays in Fig.~\ref{Fig:ExclusionPlot2}. 
On the left-hand side, we display the region of the $\alpha_{B}$-$m_{B}$ plane excluded by the BESIII Collaboration $\eta^{\prime}\to\pi^{0}\gamma\gamma$ measurement, $\rm{BR}=(3.20\pm0.07\pm0.23)\times10^{-3}$ \cite{BESIII:2016oet}, and, on the right-hand side, the corresponding one for the $\eta^{\prime}\to\eta\gamma\gamma$, $\rm{BR}=(8.25\pm3.41\pm0.72)\times10^{-5}$ \cite{BESIII:2019ofm}, both at a confidence level of $2\sigma$.

The shape of the excluded region for the $\eta^{\prime}\to\pi^{0}\gamma\gamma$ is clearly different to that of the $\eta\to\pi^{0}\gamma\gamma$ decay (cf.~Fig.~\ref{Fig:ExclusionPlot}). 
In particular, the limits within the $m_{\pi^{0}}\lesssim m_{B}\lesssim m_{\eta}$ mass range, whilst still showing the shape resembling a keel, are about four orders of magnitude weaker than those coming from $\eta\to\pi^{0}\gamma\gamma$. There are two contributing effects required to explain this. On the one hand, there are inherent dynamical differences in the $B$-boson production of the two decays [cf.~Eq.~(\ref{Eq:DecayWidthPtoBgamma})].
On the other hand, there are kinematic influences that also need to be accounted for. Specifically, if one applies the NWA to both the $B$ boson\footnote{This approximation is reasonable for $m_B\lesssim 600$ MeV, as can be checked in Fig.~\ref{Fig:BbosonTotalWidth}. Beyond this point, the use of the NWA may be questionable. As we are only attempting to provide a qualitative explanation, this limitation does not really concern us here.} and $\omega$ propagators, a factor like $\delta(t-m_B^2)\delta(t-m_{\omega}^2)$ is obtained, which under the phase-space integral results in $\delta(m_{\omega}^2-m_B^2)$.\footnote{There is no need to consider the $\phi$ propagator given that the available phase space does not allow the $\phi$ to resonate.} This Dirac delta suppresses 
the contribution of the $B$ boson to the decay process when $m_B\not\approx m_{\omega}$ and, hence, forces the exclusion limit to be weaker in this region. 
In contrast, the $B$-boson contribution is largely amplified when $m_B\approx m_{\omega}$ and, therefore, the exclusion limit becomes much stronger in this area, which will make it difficult to experimentally identify a $B$ boson with a mass around the pole of the $\omega$ resonance.
The region $m_{\eta}\lesssim m_{B}\lesssim m_{\omega}$ is less constrained, and, thus, appears to be a good place to look for an enhancement in the $\pi^{0}\gamma$ invariant mass spectrum.

The limits from the $\eta^{\prime}\to\eta\gamma\gamma$ process (right plot in Fig.~\ref{Fig:ExclusionPlot2}) in the $m_{B}\gtrsim m_{\eta}$ region are similar to the ones from $\eta^{\prime}\to\pi^{0}\gamma\gamma$. Having said that, the keel shape appearing in the $m_{\pi^{0}}\lesssim m_{B}\lesssim m_{\eta}$ mass range of the $\eta\to\pi^{0}\gamma\gamma$ and $\eta^{\prime}\to\pi^{0}\gamma\gamma$ exclusion plots is missing in the $\eta^{\prime}\to\eta\gamma\gamma$ one, which is down to the fact that the phase space of the latter does not allow the $B$ boson to resonate in this range of $B$-boson masses and, therefore, the constraints turn out to be weaker.

All in all, the $\eta^{\prime}\to\pi^{0}\gamma\gamma$ and $\eta^{\prime}\to\eta\gamma\gamma$ decays do not appear to be as powerful as the $\eta\to\pi^{0}\gamma\gamma$ for constraining the $B$-boson parameters.

\begin{figure}
\centering\includegraphics[scale=0.375]{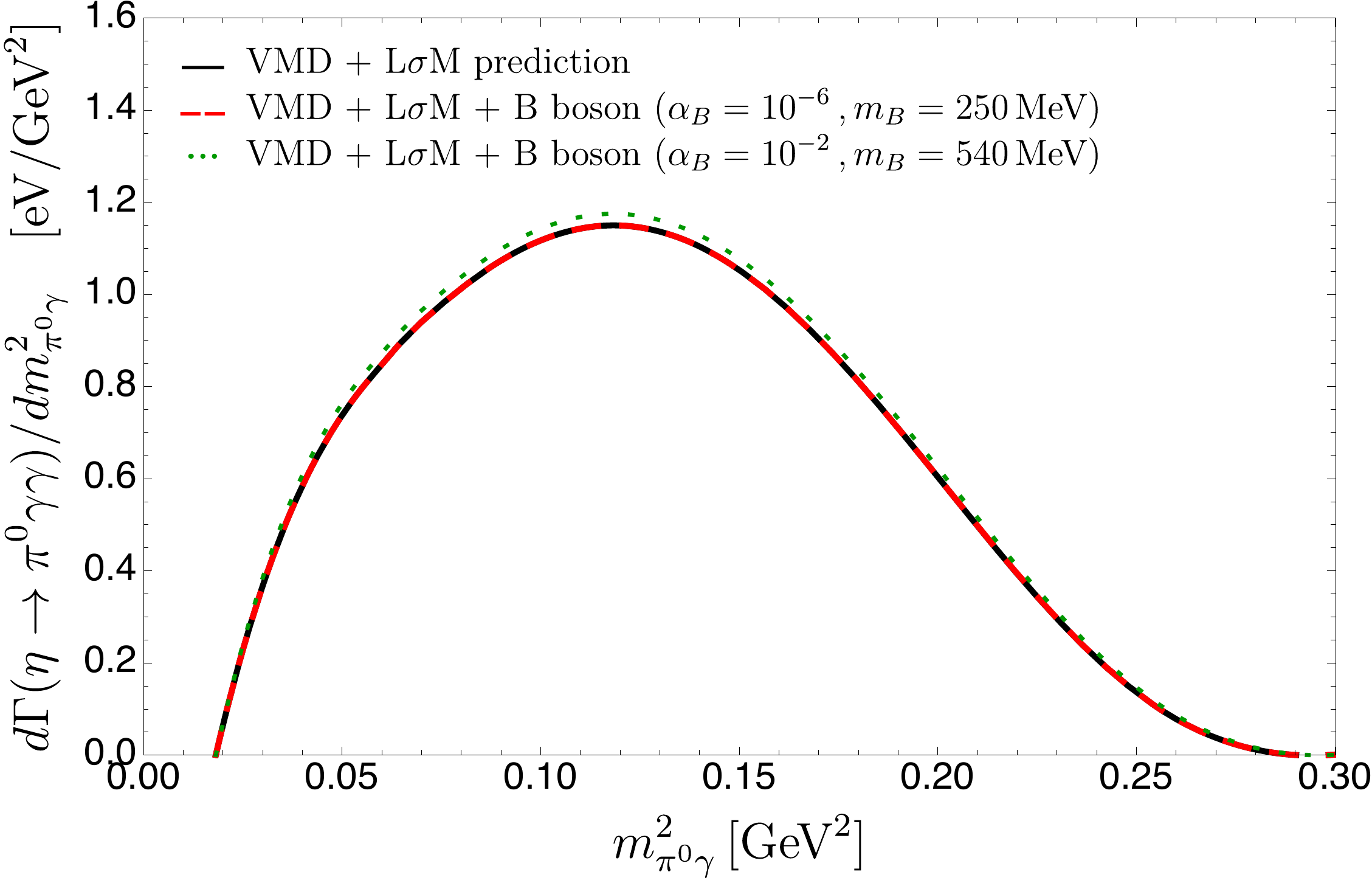}
\caption{$m_{\pi^{0}\gamma}^{2}$ distribution for the $\eta\to\pi^{0}\gamma\gamma$ decay using our theoretical VMD and L$\sigma$M prediction~\cite{Escribano:2018cwg} (solid black line). Also shown are the spectra including the $B$-boson contribution using the two sets of representative values for $\alpha_{B}$ and $m_{B}$ from Eqs.~(\ref{Eq:Set1}) (dashed red line) and (\ref{Eq:Set2}) (dotted green line).}
\label{Fig:mpi0gammaSpectrum} 
\end{figure}

The smoking gun signature of a $B$ boson in the $m_{\pi^{0}}\lesssim m_{B}\lesssim m_{\eta}$ region would be the observation of a peak at around $m_{B}$ in the $\pi^{0}\gamma$ invariant mass distribution.
In Fig.~\ref{Fig:mpi0gammaSpectrum}, we show the quantitative effect of a $B$ boson on the $\eta\to\pi^0\gamma\gamma$ decay using two sets of representative values for $\alpha_{B}$ and $m_{B}$ from the not-excluded region of parameter space
\begin{equation}
  \alpha_{B}=10^{-6},\quad m_{B}=250\,\rm{MeV}\ ,  
  \label{Eq:Set1}
\end{equation}
and
\begin{equation}
  \alpha_{B}=10^{-2},\quad m_{B}=540\,\rm{MeV}\ .
  \label{Eq:Set2}
\end{equation}
In this figure, the solid black line corresponds to our SM prediction from Ref.~\cite{Escribano:2018cwg}, whereas the effect of including the $B$ boson is shown by the dashed red and dotted green lines for the two sets of $\alpha_{B}$ and $m_{B}$ values from Eqs.~(\ref{Eq:Set1}) and (\ref{Eq:Set2}), respectively.
As it can be seen, the differences in the distribution introduced by the $B$-boson contribution are very small and it is very difficult to distinguish the associated lines from the SM prediction.
That is, the allowed values for $\alpha_{B}$ in the $m_{\pi^{0}}\lesssim m_{B}\lesssim m_{\eta}$ region are so small that it makes the $B$-boson signal strongly suppressed, rendering the task of experimentally identifying it nearly impossible.
For this reason, a $B$ boson in the mass range $m_{\pi^{0}}\lesssim m_{B}\lesssim m_{\eta}$ cannot explain the normalization offset that appears to be affecting the experimental $\gamma\gamma$ invariant mass distribution from the A2~\cite{Nefkens:2014zlt} and Crystal Ball~\cite{Prakhov:2008zz} Collaborations with respect to our VMD and L$\sigma$M prediction~\cite{Escribano:2018cwg}.

\begin{figure}
\centering\includegraphics[scale=0.375]{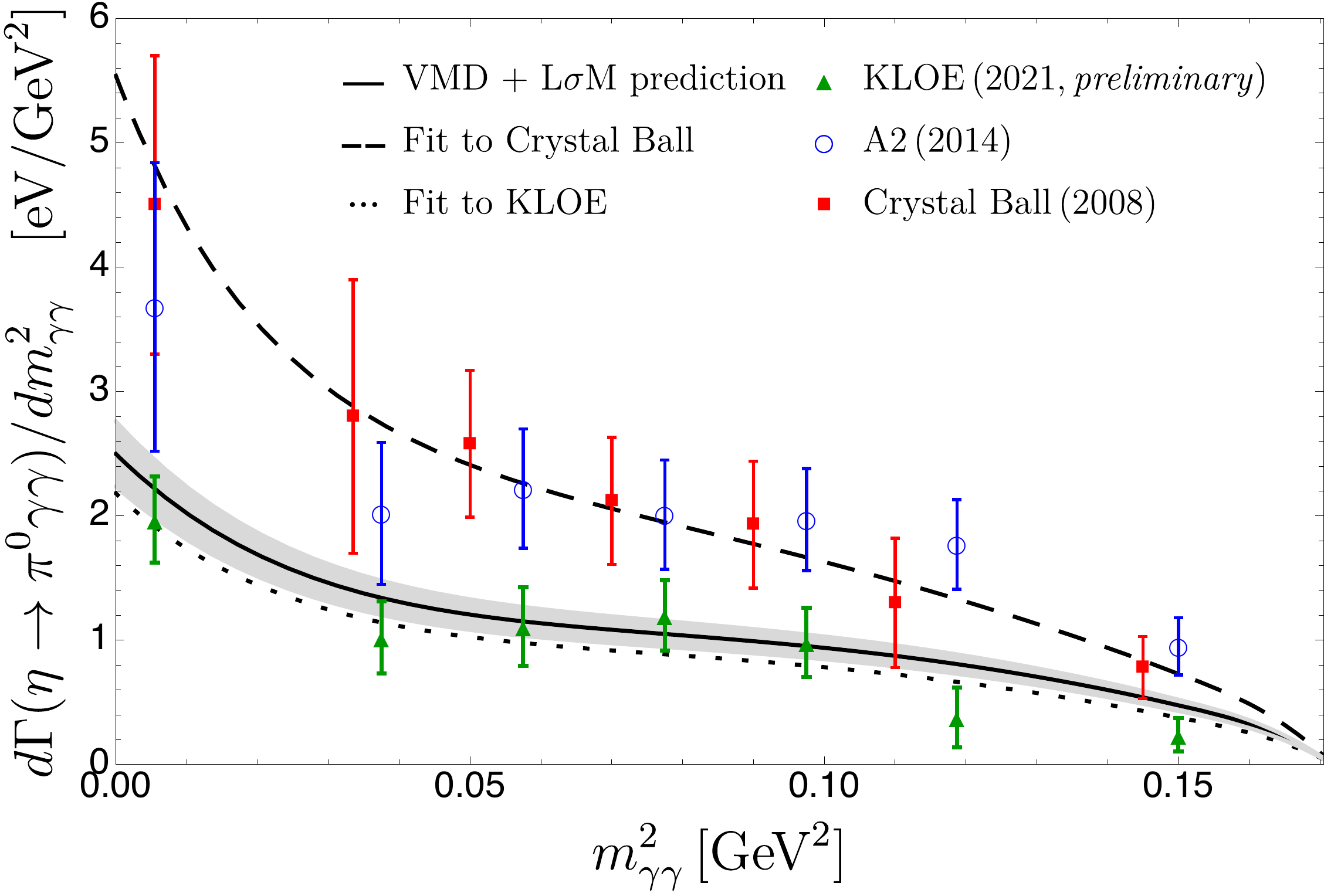}
\caption{KLOE (green triangles)~\cite{KLOECD2021}, A2 (blue circles)~\cite{Nefkens:2014zlt} and 
Crystal Ball (red squares)~\cite{Prakhov:2008zz} measurements of the $m_{\gamma\gamma}^{2}$ spectrum for the 
$\eta\to\pi^{0}\gamma\gamma$ decay, as well as the SM (VMD and L$\sigma$M) prediction~\cite{Escribano:2018cwg} (solid black line) 
and SM with $B$-boson predictions using the fitted parameters from Eqs.~(\ref{Eq:FitResultsCB}) and (\ref{Eq:FitResultsKLOE}).
}
\label{Fig:FitResultsEta} 
\end{figure}

Let us now move on to perform statistical fits to the available experimental diphoton spectra to determine the region of the $\alpha_{B}$-$m_{B}$ plane (cf.~Fig.~\ref{Fig:ExclusionPlot}) that is preferred by the data.
From the Crystal Ball $\gamma\gamma$ invariant mass spectrum~\cite{Prakhov:2008zz}, we obtain the following best fit values
\begin{equation}
\alpha_{B}=0.40^{+0.07}_{-0.08}\,,\quad m_{B}=583^{+32}_{-20}\,\,\rm{MeV}\,,    
\label{Eq:FitResultsCB}
\end{equation}
with a $\chi^{2}_{\rm{min}}\rm{/d.o.f}=0.42/5=0.08$, whereas, for the KLOE (preliminary) data~\cite{KLOECD2021}\footnote{Whilst KLOE has published a BR for the $\eta\to\pi^{0}\gamma\gamma$ process in a conference proceedings~\cite{Cao:2022rxo}, the diphoton spectrum has not yet been published, although it was presented at The 10th International Workshop on Chiral Dynamics 2021~\cite{KLOECD2021}. For our analysis, we have retrieved the data points from their presentation's figure. We thank KLOE for the email communications~\cite{KLOECD2021privatecorrespondence}.}, we find
\begin{equation}
\alpha_{B}=0.049^{+40}_{-27}\,,\quad m_{B}=135^{+1}_{-135}\,\,\rm{MeV}\,,     
\label{Eq:FitResultsKLOE}
\end{equation}
with a $\chi^{2}_{\rm{min}}\rm{/d.o.f}=4.46/5=0.89$.\footnote{We also carried out fits to the A2 data but did not find convergent solutions using two free parameters. When fits were attempted using the $B$-boson width as an additional free parameter, though, good convergence was achieved.} 
Because of the large errors associated to the experimental points from Crystal Ball, its $\chi^{2}_{\rm{min}}$/d.o.f turns out to be extremely small. The $\chi^{2}_{\rm{min}}$/d.o.f of the fit to the KLOE data implies a good quality of the fit.
The errors associated to the fitted parameters have been estimated by perturbing one of the parameters at a time such that $\chi^2=\chi^2_{\rm{min}}+1$~\cite{Zyla:2020zbs}.
The theoretical $\gamma\gamma$ invariant mass spectra using the parameters from the fits in Eqs.~(\ref{Eq:FitResultsCB}) and (\ref{Eq:FitResultsKLOE}) to the Crystal Ball and KLOE data are shown in Fig.~\ref{Fig:FitResultsEta} with dashed and dotted black lines, respectively. Also plotted are the experimental data points and the SM 
prediction \cite{Escribano:2018cwg} (solid black line) 
with an estimation of the uncertainty from the error propagation of the $VP\gamma$ couplings.
The different individual contributions to the invariant mass spectra are shown in Appendix~\ref{Appendix11}.
It is worth noticing that the inclusion of a nonresonant $B$ boson in the $t$ and $u$ channels, with parameters from Eq.~(\ref{Eq:FitResultsCB}), helps explain the tension between the Crystal Ball spectrum and the SM result \cite{Escribano:2018cwg}. 
Notwithstanding this, the best fit parameters from Crystal Ball in Eq.~(\ref{Eq:FitResultsCB}) are ruled out by the KLOE data (cf.~Fig.~\ref{Fig:ExclusionPlot}), whose measured BR continues the decreasing trend seen over the decades associated to more precise measurements becoming available (see Ref.~\cite{Achasov:2001qm}).
In turn, this trend supports the theoretical treatment without a $B$ boson, as our VMD and L$\sigma$M approach from Ref.\,\cite{Escribano:2018cwg} appears to be capable of successfully predicting the experimental data for the three $\eta^{(\prime)}\to\pi^{0}\gamma\gamma$ and $\eta^{\prime}\to\eta\gamma\gamma$ decays simultaneously.
Clearly, the experimental situation is far from conclusive and it may not be possible to make categorical statements about the need for a $B$ boson until the arrival of new and more precise data, e.g.~from the KLOE(-II) and JEF~\cite{JEF} experiments.

\begin{figure}
\centering\includegraphics[scale=0.375]{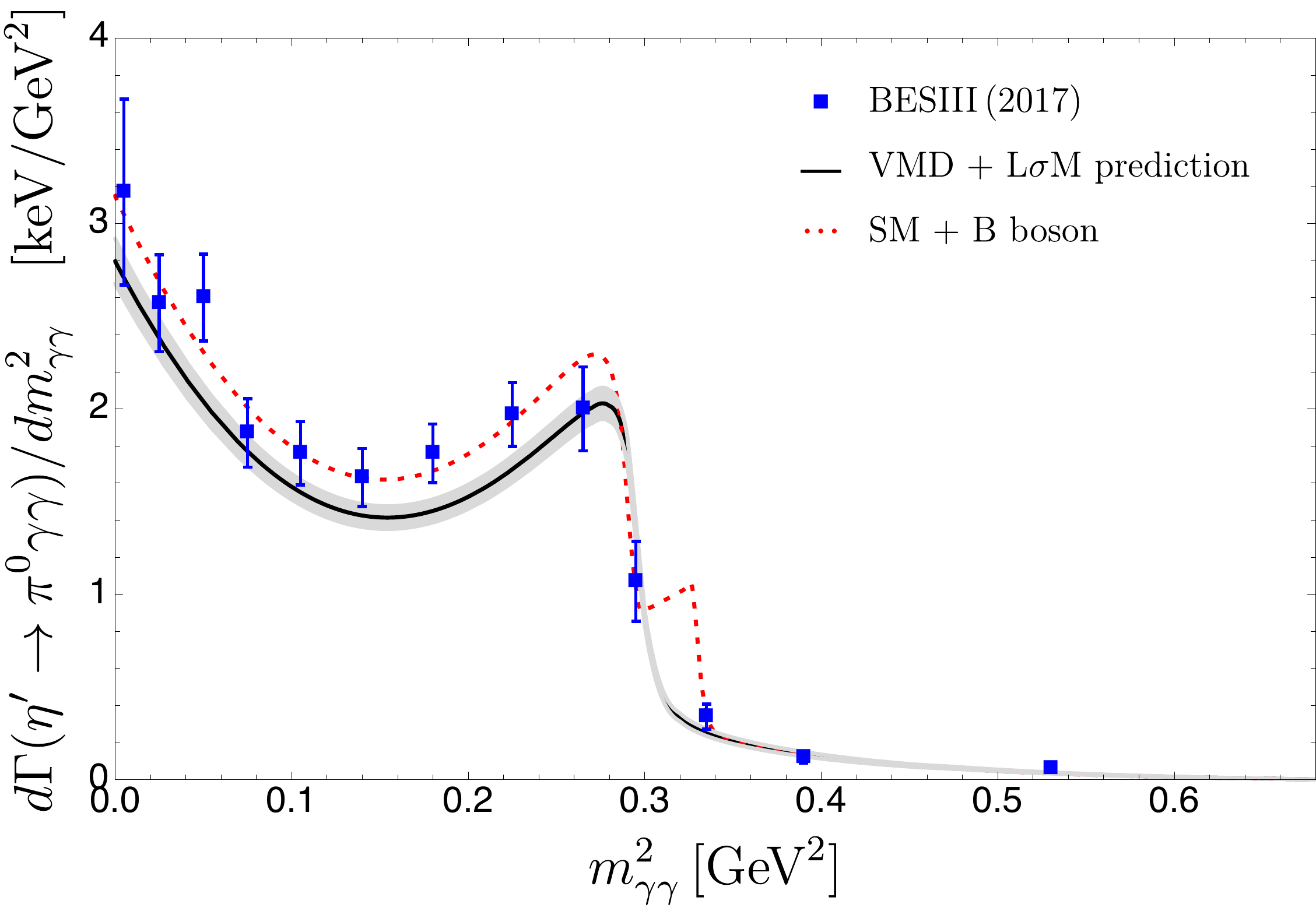}
\caption{BESIII (blue squares)~\cite{BESIII:2016oet} measurements of the $m_{\gamma\gamma}^{2}$ spectrum
for the $\eta^{\prime}\to\pi^{0}\gamma\gamma$ decay, 
as well as the SM (VMD and L$\sigma$M) prediction~\cite{Escribano:2018cwg} (solid black line) 
and SM with $B$-boson prediction using the fitted parameters from Eq.~(\ref{Eq:FitResultsBESIII1}) (dotted red line).}
\label{Fig:FitResultsEtaP} 
\end{figure}

Next, we perform fits to the $\eta^{\prime}\to\pi^{0}\gamma\gamma$ diphoton spectrum from the 
BESIII Collaboration~\cite{BESIII:2016oet}, which may be used to explore larger $B$-boson masses. 
No distribution data is available for the $\eta^{\prime}\to\eta\gamma\gamma$ process, 
so the constraints from this channel come from the branching ratio only (see Fig.~\ref{Fig:ExclusionPlot2}).
The fit to the $\eta^{\prime}\to\pi^{0}\gamma\gamma$ data yields 
\begin{equation}
\alpha_{B}=0.005(1)\,,\quad m_{B}=759(1)\,\,\rm{MeV}\,, 
\label{Eq:FitResultsBESIII1}
\end{equation}
with $\chi^{2}_{\rm{min}}\rm{/d.o.f}=11.73/11=1.07$. 
The distribution using the fitted parameters from Eq.~(\ref{Eq:FitResultsBESIII1}) is shown in Fig.~\ref{Fig:FitResultsEtaP} (dotted red line), together with the experimental data (blue squares) and the SM prediction~\cite{Escribano:2018cwg} (solid black line) with an estimation of its uncertainty. 
It is worth noticing the sudden drop in the dotted red line (i.e.~SM with $B$-boson distribution) at $m_{\gamma\gamma}^2\approx 0.33$ GeV$^2$.\footnote{This sudden drop can be understood as follows: so long as $t_{\rm{min}}(s)\leq m_B^2 \leq t_{\rm{max}}(s) \ \forall \ s \in [s_{\rm{min}},s_{\rm{max}}]$, then the available phase space allows the $B$ boson to resonate; however, for values of $s$ such that $t_{\rm{min}}(s)\geq m_B^2$ or $t_{\rm{max}}(s)\leq m_B^2$, then the $B$ boson no longer resonates and its contribution to the amplitude suddenly plummets producing the sudden drop in the distribution. 
It must be noted that this effect also applies to the $\omega$ meson and is responsible for the sudden drop in the SM distribution around $m^2_{\gamma\gamma}\approx 0.30$ GeV$^2$. Given that this is a kinematic effect, it will always be present in the spectrum so long as $t_{\rm{min}}(s_{\rm{min}})\leq m_B^2\leq t_{\rm{max}}(s_{\rm{min}})$, although it becomes a relatively small effect and is difficult to detect beyond $m^2_{\gamma\gamma}\gtrsim0.4$ GeV$^2$ for this particular decay.} What is interesting about this is that, even though the $\chi^2_{\rm{min}}\rm{/d.o.f}$ of this fit is very good, the associated integrated branching ratio deviate from the experimental counterpart due to the effect of the wiggle on the distribution. Also, the spectrum using the fit parameters would lead to larger than observed bin values for the experimental points 10 and 11. 
Again, the different individual contributions to the ${\gamma\gamma}$ invariant mass spectrum are presented in Appendix~\ref{Appendix12}.

\begin{figure}
\includegraphics[scale=0.375]{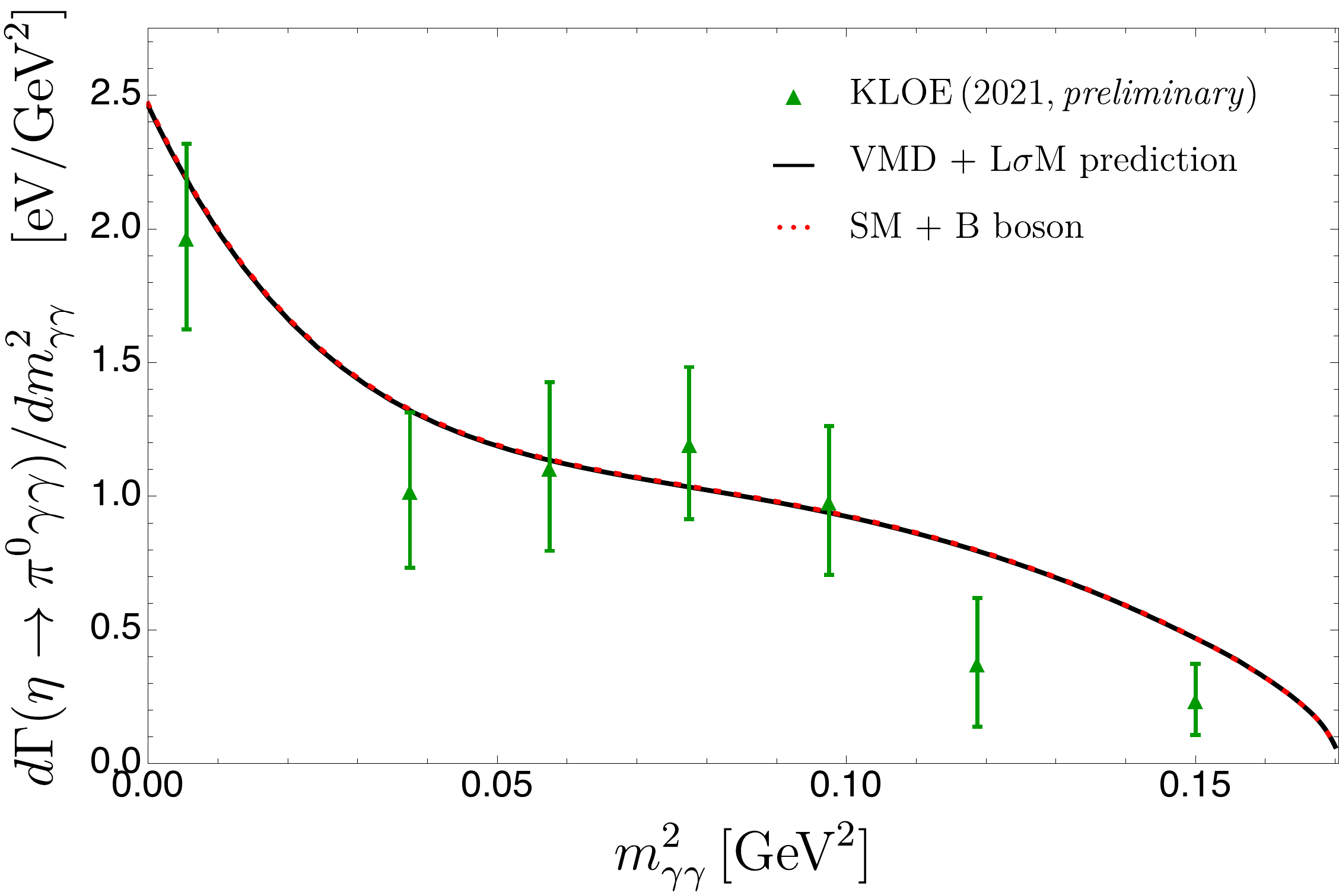} 
\caption{KLOE measurements (green triangles)~\cite{KLOECD2021} of the $m_{\gamma\gamma}^{2}$ spectrum
for the $\eta\to\pi^{0}\gamma\gamma$ decay 
together with the SM (VMD and L$\sigma$M) prediction~\cite{Escribano:2018cwg} (solid black line)
and SM with $B$-boson prediction using the fitted parameters from Eq.~(\ref{Eq:FitResultsKLOEBESIII1}) (dotted red line).}
\label{Fig:FitResultsKLOEBESIII} 
\end{figure}

Finally, a joint fit to the experimental invariant mass spectra from KLOE and BESIII for the $\eta\to\pi^0\gamma\gamma$ and $\eta^{\prime}\to\pi^0\gamma\gamma$ decays, respectively, is carried out. The joint fit yields
\begin{equation}
\alpha_{B}=0.005(1)\,,\quad m_{B}=759(1)\,\,\rm{MeV}\,, 
\label{Eq:FitResultsKLOEBESIII1}
\end{equation}
with $\chi^{2}_{\rm{min}}\rm{/d.o.f}=19.61/18=1.09$.\footnote{There is a secondary local minimum giving $\alpha_{B}=5(2)\times 10^{-4}$ and $m_{B}=780^{+3}_{-4}\,\,\rm{MeV}$, with $\chi^{2}_{\rm{min}}\rm{/d.o.f}=23.71/18=1.32$. These values for the $\alpha_B$ and $m_B$ parameters yield a $B$-boson width of $\Gamma_B\approx5.1$ MeV. For this particular solution, both $m_B$ and $\Gamma_B$ are effectively the same as those of the $\omega$ vector meson. Accordingly, the end effect is to enhance the spectrum  with respect to the SM prediction for $m_{\gamma\gamma}^2\lesssim 0.30$ GeV$^2$, where the available phase space allows both the $\omega$ and $B$ boson to resonate, and has no effect on the spectrum beyond this point, i.e.~$m_{\gamma\gamma}^2\gtrsim$ 0.30 GeV$^2$.}
The joint fit produces the same best fit parameters as those from the fit to the $\eta^\prime\to\pi^0\gamma\gamma$ spectrum only. 
The theoretical distribution for the $\eta\to\pi^0\gamma\gamma$ decay using the parameters from the joint fit is shown in Fig.~\ref{Fig:FitResultsKLOEBESIII} (dotted red line), which turns out to be indistinguishable from that of the SM (solid black line). 

To conclude, it is worth highlighting that both our SM and SM with $B$-boson predictions (using the joint fit parameters for the latter) agree well with both sets of experimental data points. The largest differences between the theoretical predictions still show compatibility at roughly the $1\sigma$ level. We, therefore, conclude that the experimental data from KLOE and BESIII for the $\eta\to\pi^0\gamma\gamma$ and $\eta^{\prime}\to\pi^0\gamma\gamma$ decays, respectively, do not require a $B$-boson contribution, in spite of the coupling $\alpha_B$ being clearly nonzero.

This conclusion differs from that of the study in Ref.~\cite{Balytskyi:2021lzh}, where it is argued that the simultaneous prediction of the three processes under study may require the presence of a leptophobic $B$ boson, which, in turn, was motivated by some of the conclusions from our previous work in Ref.~\cite{Escribano:2018cwg}.
It should be noted, though, that in Ref.~\cite{Balytskyi:2021lzh} the $B$-boson mass and width were manually fixed to some values that the author deemed reasonable, 
leaving the $g_{BP\gamma}$ couplings as free constant parameters that were subsequently fitted to the experimental data. Contrary to this, in the present work $\Gamma_B$ is not an independent variable but a function of both $\alpha_B$ and $m_B$ (under the assumption that the $B$ boson decays to SM particles only), which are then left as free parameters in our fits. 
It is worth noting that in our analysis the $g_{BP\gamma}$ couplings are not constant but energy dependent.
More importantly, Ref.~\cite{Balytskyi:2021lzh} did not employ the most recent experimental data for the $\eta\to\pi^0\gamma\gamma$ decay from the KLOE Collaboration~\cite{KLOECD2021} in his analysis. 



\section{Conclusions}\label{sectionConclusions}

\begin{figure}
\centering\includegraphics[scale=0.375]{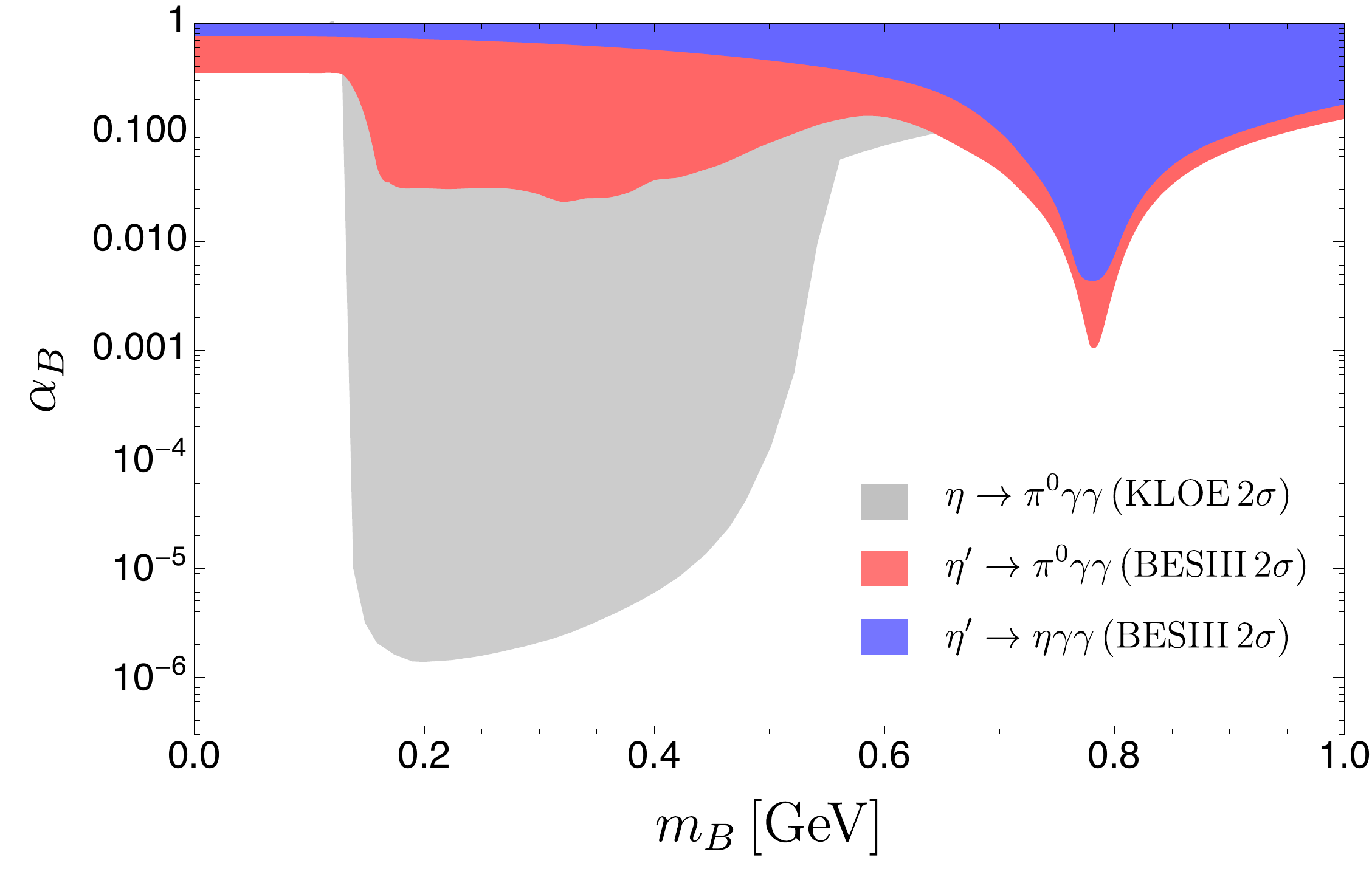}
\caption{Limits on the leptophobic $B$-boson mass $m_{B}$ and coupling $\alpha_{B}$ 
from the BR measurements of the decays $\eta\to\pi^{0}\gamma\gamma$ (gray) by KLOE~\cite{Cao:2022rxo}, 
and $\eta^{\prime}\to\pi^{0}\gamma\gamma$ (red)~\cite{BESIII:2016oet} 
and $\eta^{\prime}\to\eta\gamma\gamma$ (blue)~\cite{BESIII:2019ofm} by BESIII.}
\label{Fig:ExclusionPlotTotal} 
\end{figure}

We have analyzed in detail the sensitivity of the rare decays 
$\eta^{(\prime)}\to\pi^{0}\gamma\gamma$ and $\eta^{\prime}\to\eta\gamma\gamma$ 
to a leptophobic $U(1)_{B}$ boson in the MeV--GeV mass range.
Adding the explicit $B$-boson exchange contribution in the $t$ and $u$ channels, 
in addition to our SM (VMD and L$\sigma$M) amplitudes, 
has allowed us to place stringent limits on the $B$-boson parameters $m_{B}$ and $\alpha_{B}$ 
by comparing with current experimental data.
A visual summary of these limits is shown in Fig.~\ref{Fig:ExclusionPlotTotal}.
From the individual analysis of the $\eta\to\pi^{0}\gamma\gamma$ decay, 
we have strengthened by one order of magnitude the current constraints in the resonant mass region 
$m_{\pi^{0}}\lesssim m_{B}\lesssim m_{\eta}$, reaching $\alpha_{B}\sim10^{-6}$, as it can be seen in the figure.
These constraints would make a $B$-boson signature strongly suppressed, 
rendering the task of experimentally identifying it as a peak around $m_B$ in the ${\pi^0\gamma}$ 
invariant mass distribution practically impossible.

Our analysis of the most recent experimental ${\gamma\gamma}$ invariant mass distribution from the KLOE Collaboration 
supports the description of the processes studied in this work without contribution from a potential new leptophobic $B$ boson, 
as our VMD and L$\sigma$M treatment is capable of simultaneously predicting the three 
$\eta^{(\prime)}\to\pi^{0}\gamma\gamma$ and $\eta^{\prime}\to\eta\gamma\gamma$ decays 
with remarkable agreement with the experimental data.
However, a $B$ boson with a mass $m_B\gtrsim m_{\eta}$ and non-negligible coupling $\alpha_B$ may help explain the discrepancy 
between our SM prediction and the experimental data from the A2 and Crystal Ball Collaborations (see Fig.~\ref{Fig:FitResultsEta}).
The existing tension between the measurements by different experimental groups does not allow us 
to make an absolute statement about the need for a $B$ boson, 
as the branching ratio observed by KLOE, whilst in agreement with our SM prediction, 
is about a factor of two smaller than those from A2 and Crystal Ball.
This highlights the need for new and more precise data,~e.g.~from the KLOE(-II) and JEF experiments.

Finally, the $\eta^{\prime}\to\pi^{0}\gamma\gamma$ and $\eta^{\prime}\to\eta\gamma\gamma$ decays are not as powerful as the 
$\eta\to\pi^{0}\gamma\gamma$ at constraining $B$-boson parameters below $m_{\eta}$ but allow exploring larger $B$-boson masses.
As it can be observed in Fig.~\ref{Fig:ExclusionPlotTotal}, 
the region in the $\alpha_{B}$-$m_{B}$ plane near the $\omega$ pole shows a sharp dip, 
which would make the task of identifying a $B$ boson with $m_{B}\sim m_{\omega}$ very challenging. 
However, the mass region $m_{\eta}\lesssim m_{B}\lesssim m_{\omega}$ is less constrained, and, thus, 
appears to be a good place to look for an enhancement in the ${\pi^{0}(\eta)\gamma}$ invariant mass distributions,
for example, at BESIII or the JEF experiment.


\begin{acknowledgements}

The work of S.~G-S.~is supported by the Laboratory Directed Research and Development program of Los Alamos National Laboratory 
under Project No.~20210944PRD2, and by the U.S.~Department of Energy through the Los Alamos National Laboratory.
Los Alamos National Laboratory is operated by Triad National Security, LLC, for the National Nuclear Security Administration of 
U.S.~Department of Energy (Contract No.~89233218CNA000001).
The work of R.~E.~and E.~R.~has been supported by 
the Spanish Ministry of Science and Innovation under Project No.~PID2020-112965GB-I00, 
and by the European Union’s Horizon 2020 Research and Innovation Programme under grant 
no.~824093 (H2020- INFRAIA-2018-1). 
IFAE is partially funded by the CERCA program of the Generalitat de Catalunya.
\end{acknowledgements}

\appendix


\section{Model uncertainty analysis}\label{Appendix2}

\begin{figure}
\centering\includegraphics[scale=0.375]{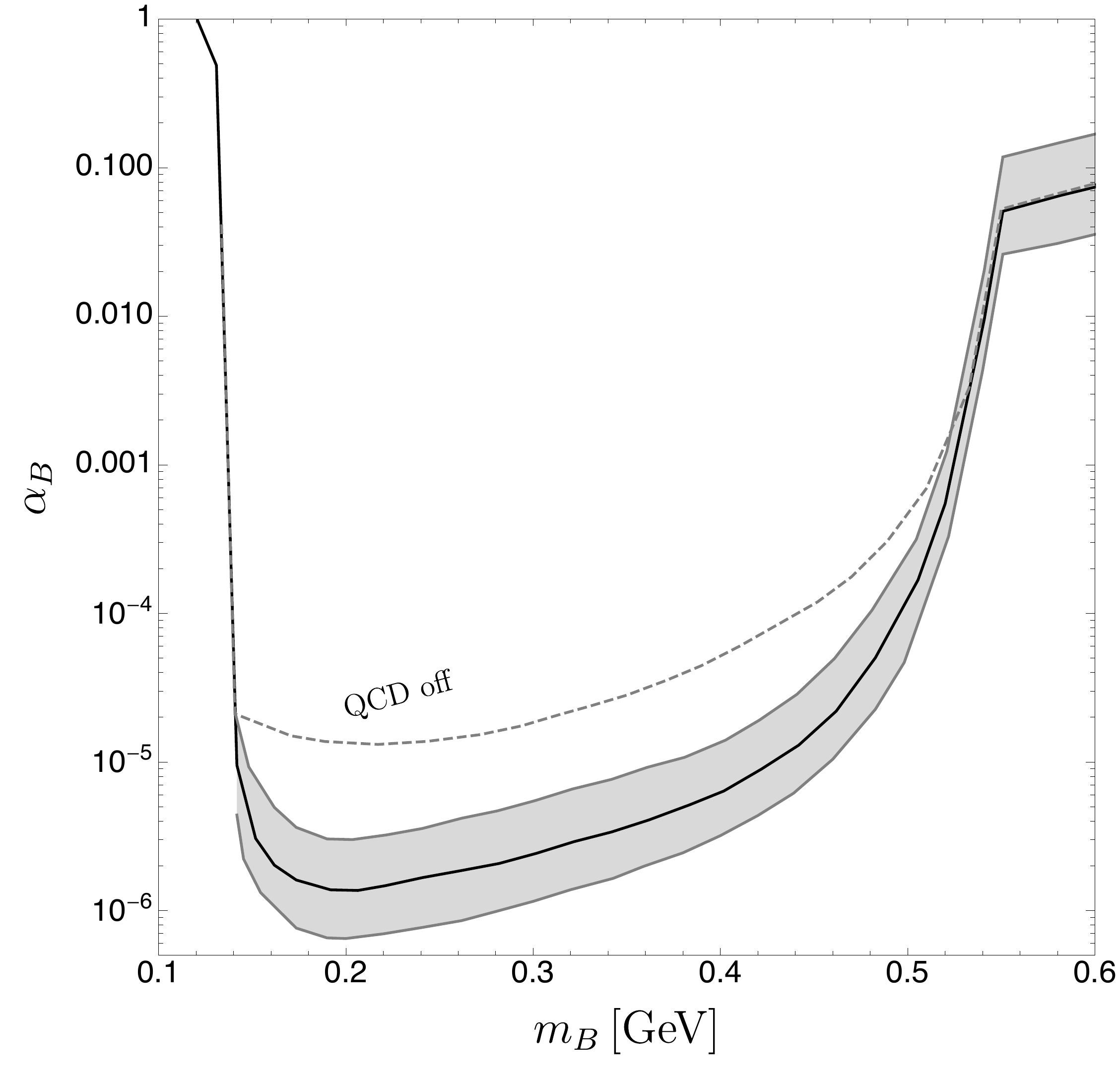}
\caption{Limits on the leptophobic $B$-boson coupling $\alpha_{B}$ for different $m_{B}$ masses 
from the $\eta\to\pi^{0}\gamma\gamma$ BR measurement by KLOE~\cite{Cao:2022rxo} (black line) 
along with an error band due to the uncertainties of the $VP\gamma$ couplings 
in Table \ref{gVPgammacouplingsempirical}.}
\label{Fig:ExclusionPlotError} 
\end{figure}

In this appendix, we provide an estimation of the uncertainty related
to the models employed in our theoretical treatment. 
In particular, 
and based on the arguments laid out in the last two paragraphs of Sec.~\ref{subsection21}, 
one can assess this uncertainty 
by just propagating the errors associated to the couplings in 
Table \ref{gVPgammacouplingsempirical} to the final results.
In Fig.~\ref{Fig:ExclusionPlotError}, 
we show the limits on the leptophobic $B$-boson coupling $\alpha_{B}$ for different $m_{B}$ masses 
from the $\eta\to\pi^{0}\gamma\gamma$ BR measurement by KLOE (black line) along with an error band 
due to the uncertainties of the $VP\gamma$ couplings (gray band). 
Despite the error band being rather large, 
the corresponding limits are clearly different from those with the QCD contributions turned off 
(dashed line).

\section{Individual contributions to the invariant mass distributions}\label{Appendix1}

\subsection{$\eta\to\pi^{0}\gamma\gamma$ decay}\label{Appendix11}
\begin{figure*}
\centering\includegraphics[scale=0.375]{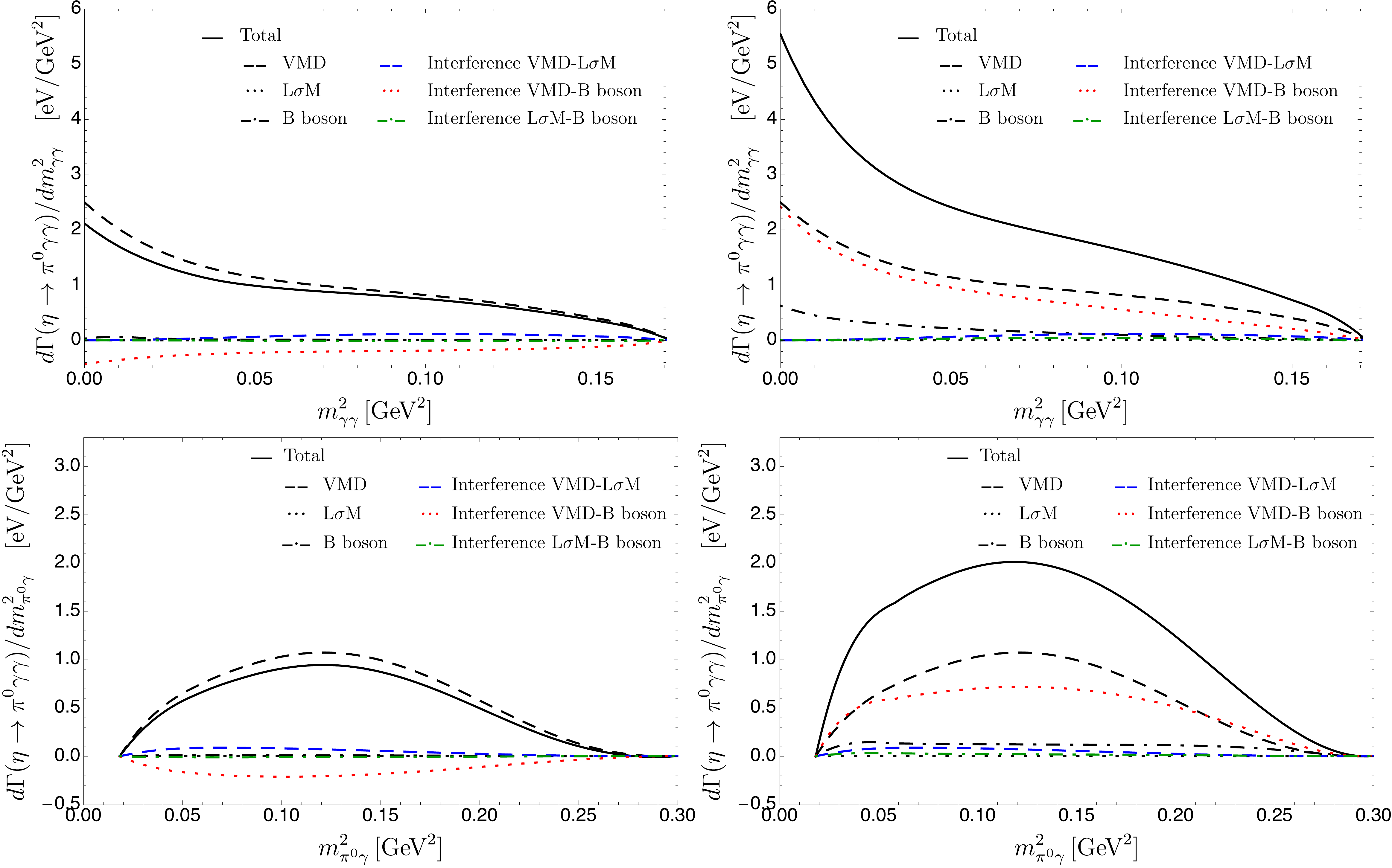}
\caption{Different individual contributions to the 
$m_{\gamma\gamma}^{2}$ (first row) and $m_{\pi^{0}\gamma}^{2}$ (second row) 
distributions for the $\eta\to\pi^{0}\gamma\gamma$ decay 
corresponding to the fit results 
shown in Eqs.~(\ref{Eq:FitResultsKLOE}) and (\ref{Eq:FitResultsCB})
obtained from the KLOE (first column) and Crystal Ball (second column) experimental data, 
respectively.}
\label{Fig:IndividualContributions} 
\end{figure*}

For completeness, in Fig.~\ref{Fig:IndividualContributions},
the different individual contributions to the ${\gamma\gamma}$ (first row) 
and ${\pi^{0}\gamma}$ (second row) invariant mass distributions 
for the $\eta\to\pi^{0}\gamma\gamma$ decay are presented. 
The first and second columns in this figure correspond to the fit results 
shown in Eqs.~(\ref{Eq:FitResultsKLOE}) and (\ref{Eq:FitResultsCB})
obtained from the KLOE and Crystal Ball experimental datasets, respectively.
The entire contribution to the spectra (solid black), 
together with the separate VMD (dashed black), L$\sigma$M (dotted black) 
and $B$-boson (dot-dashed black) contributions, as well as their interferences, are displayed.

As observed, 
the exchange of vector mesons dominate the KLOE spectra, 
contributing $114\%$ to the entire signal.
The contribution of scalar exchanges accounts for less than $1\%$, 
whilst the contribution from the $B$ boson is only $1.9\%$,
making their separate effects very challenging to isolate,
even with the arrival of new and more precise data.
The interference between vector and scalar exchanges is constructive 
and accounts for about $8\%$, 
whereas the interference between vector and $B$-boson exchanges is destructive, $-23\%$,
with a visible effect in the distributions at low ${\gamma\gamma}$ and ${\pi^{0}\gamma}$
invariant masses.
The interference between scalar and $B$-boson exchanges is tiny, with less than $-1\%$,
and destructive.

The contributions to the Crystal Ball spectra show a rather different behavior.
Again, the contribution of vector meson exchanges prevails, 
but in this case it accounts only for $49\%$ of the entire signal.
The interference between vector and $B$-boson exchanges is constructive and almost as large as 
the VMD signal, contributing with $38\%$.
The remaining $13\%$ comes from the individual effect of the $B$ boson, 
which accounts for $8\%$, the interference between vector and scalar exchanges, 
accounting for $4\%$, and the interference between scalar and $B$-boson exchanges, 
which is constructive and around $1\%$.
The contribution of scalar meson exchanges is very small. 

\subsection{$\eta^{\prime}\to\pi^{0}\gamma\gamma$ decay}\label{Appendix12}

\begin{figure}
\centering\includegraphics[scale=0.375]{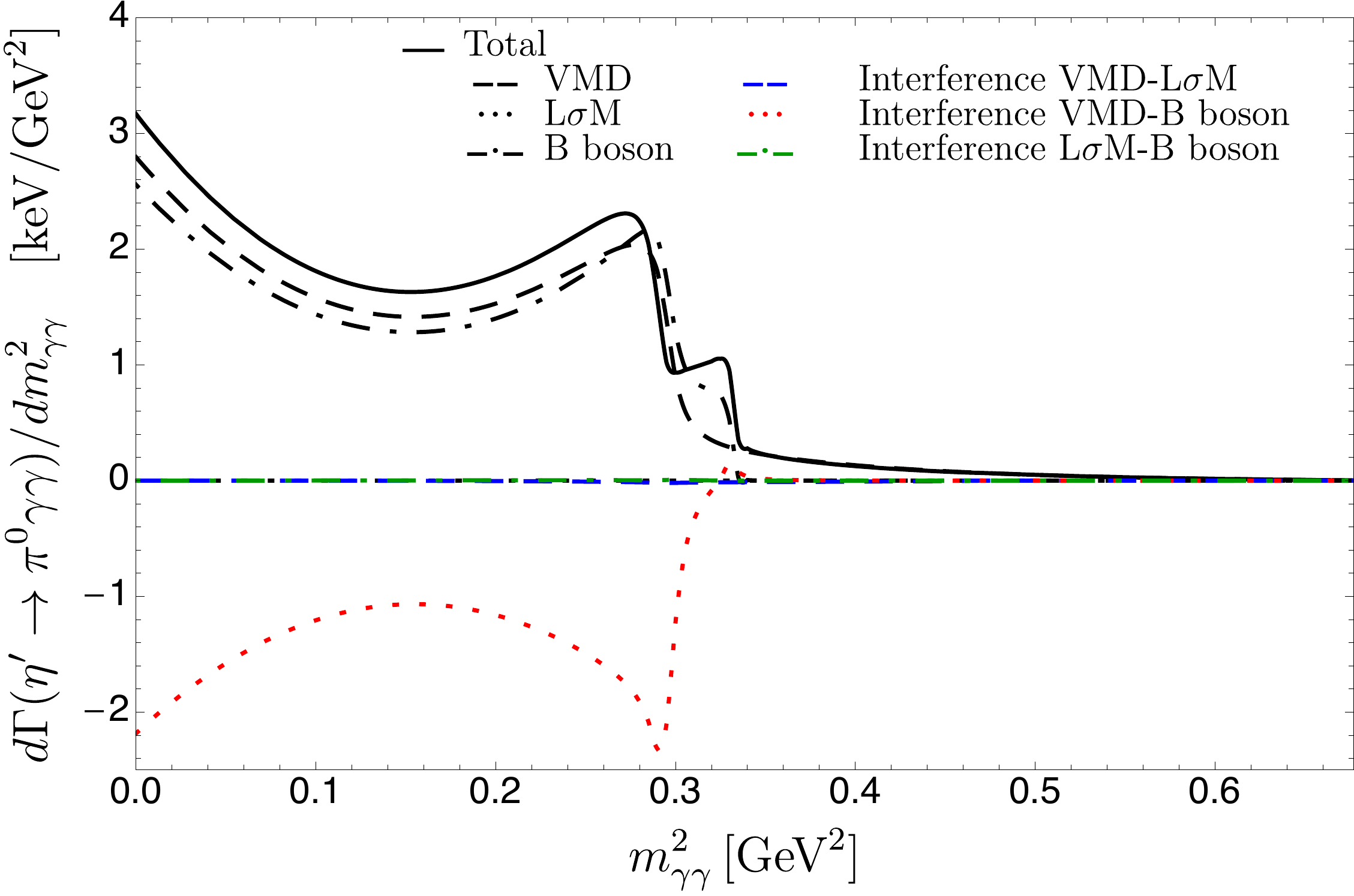}
\caption{Different individual contributions to the $m_{\gamma\gamma}^{2}$ invariant mass distribution 
for the $\eta^{\prime}\to\pi^{0}\gamma\gamma$ decay using the fit result 
in Eq.~(\ref{Eq:FitResultsBESIII1}) obtained from the BESIII experimental data.}
\label{Fig:IndividualContributionsEtaP} 
\end{figure}

In Fig.~\ref{Fig:IndividualContributionsEtaP},
the different individual contributions to the ${\gamma\gamma}$ invariant mass distribution 
for the $\eta^{\prime}\to\pi^{0}\gamma\gamma$ decay using the fit result 
in Eq.~(\ref{Eq:FitResultsBESIII1}) are also presented.
The entire contribution to the spectrum (solid black), 
together with the separate VMD (dashed black), L$\sigma$M (dotted black) 
and $B$ boson (dot-dashed black) contributions, as well as their interferences, are shown.
The exchange of vector mesons dominate, accounting for $76\%$ of the entire signal.
The individual contribution of the $B$ boson is about the same as VMD 
and the rest comes from the interference terms, 
out of which the destructive interference between vector and $B$-boson exchanges 
dominates with $-66\%$.
The contribution of scalar mesons is insignificant.

\end{document}